\documentclass{aa}  

\usepackage{graphicx}
\usepackage{txfonts}
\usepackage{color}
\begin{document} 

\title{  $Suzaku$ X-ray study of the double radio relic 
galaxy cluster\\ \object{CIZA J2242.8+5301}
  }

\author{H. Akamatsu\inst{1}, 
         R. J. van Weeren \inst{2},
         G. A. Ogrean \inst{2},
         H. Kawahara \inst{3}, 
         A. Stroe \inst{4},
         D. Sobral \inst{4, 7, 8}, \\
         M. Hoeft \inst{6}, 
         H. R\"ottgering\inst{4},
         M. Br\"uggen \inst{5},
          \and
          J. S. Kaastra \inst{1, 4}
}

\institute{SRON Netherlands Institute for Space Research, Sorbonnelaan 2, 3584 CA Utrecht, The Netherlands
                    \email{H.Akamatsu@sron.nl}
\and
Harvard-Smithsonian Center for Astrophysics, 60 Garden Street, Cambridge, MA 02138, USA
\and
Department of Earth and Planetary Science, The University of Tokyo, Tokyo 113-0033, Japan
\and
Leiden Observatory, Leiden University, PO Box 9513, 2300 RA Leiden, The Netherlands
\and
Hamburg Observatory, University of Hamburg, Gojenbergsweg 112, 21029 Hamburg, Germany
\and
Th\"uringer Landessternwarte Tautenburg, Sternwarte 5, 07778, Tautenburg, Germany
\and
Instituto de Astrof\'{\i}sica e Ci\^{e}ncias do Espa\c{c}o, Universidade de Lisboa, OAL, 
Tapada da Ajuda, PT1349-018 Lisboa, Portugal
\and
Departamento de F\'{i}sica, Faculdade de Ci\^{e}ncias, Universidade de Lisboa, Edif\'{i}cio C8, Campo Grande, PT1749-016 Lisbon, Portugal 
            }

\date{Received **; accepted **}

\abstract
{We present the results from {\it Suzaku} observations of  the merging cluster of galaxies \object{CIZA J2242.8+5301} at {\it z}=0.192.}
{To study the physics of gas heating and particle acceleration in cluster mergers, we
     investigated the X-ray emission from \object{CIZA J2242.8+5301}, which hosts  two giant radio relics in the northern/southern part of the cluster.
     }
  {We analyzed data from three-pointed {\it Suzaku} observations of  \object{CIZA J2242.8+5301} to derive the temperature distribution in four different directions. }
   { 
   The Intra-Cluster Medium (ICM) temperature shows a  remarkable drop from 
  {8.5$_{-0.6}^{+0.8}$ keV to 2.7$_{-0.4}^{+0.7}$ keV} across the northern radio relic. The temperature drop is consistent with a Mach number 
   {$ {\cal M}_n=2.7^{+0.7}_{-0.4}$ }
   and a shock velocity 
   {{$v_{shock:n}=2300_{-400}^{+700}\rm ~km~s^{-1}$}}.  
   We also confirm the temperature drop across the southern radio relic. However, the ICM temperature beyond this relic is much higher than beyond the northern one, which gives a Mach number 
   {{${\cal M}_s=1.7^{+0.4}_{-0.3}$}}
    and shock velocity 
   {{$v_{shock:s}=2040_{-410}^{+550}\rm ~km~s^{-1}$}}.  
    These results agree with other systems showing a relationship between the radio relics and shock fronts which are induced by merging activity. 
We compare the X-ray derived Mach numbers with the radio derived Mach numbers from the radio spectral index under the assumption of diffusive shock acceleration in the linear test particle regime. For the northern radio relic, the Mach numbers derived from X-ray and radio observations agree with each other.
Based on the shock velocities, we estimate that  CIZA J2242.8+5301 is observed approximately 0.6 Gyr after core passage. The magnetic field pressure at the northern relic is estimated to be 
9\%
 of the thermal pressure.
}
  {}
  \authorrunning{H. Akamatsu et al.}
  \titlerunning{$Suzaku$ X-ray observations of \object{CIZA J2242.8+5301}}

   \maketitle
%

\begin{figure*}[th]
\begin{tabular}{cc}
\begin{minipage}{0.5\hsize}
\begin{center}
\includegraphics[width=1.\hsize]{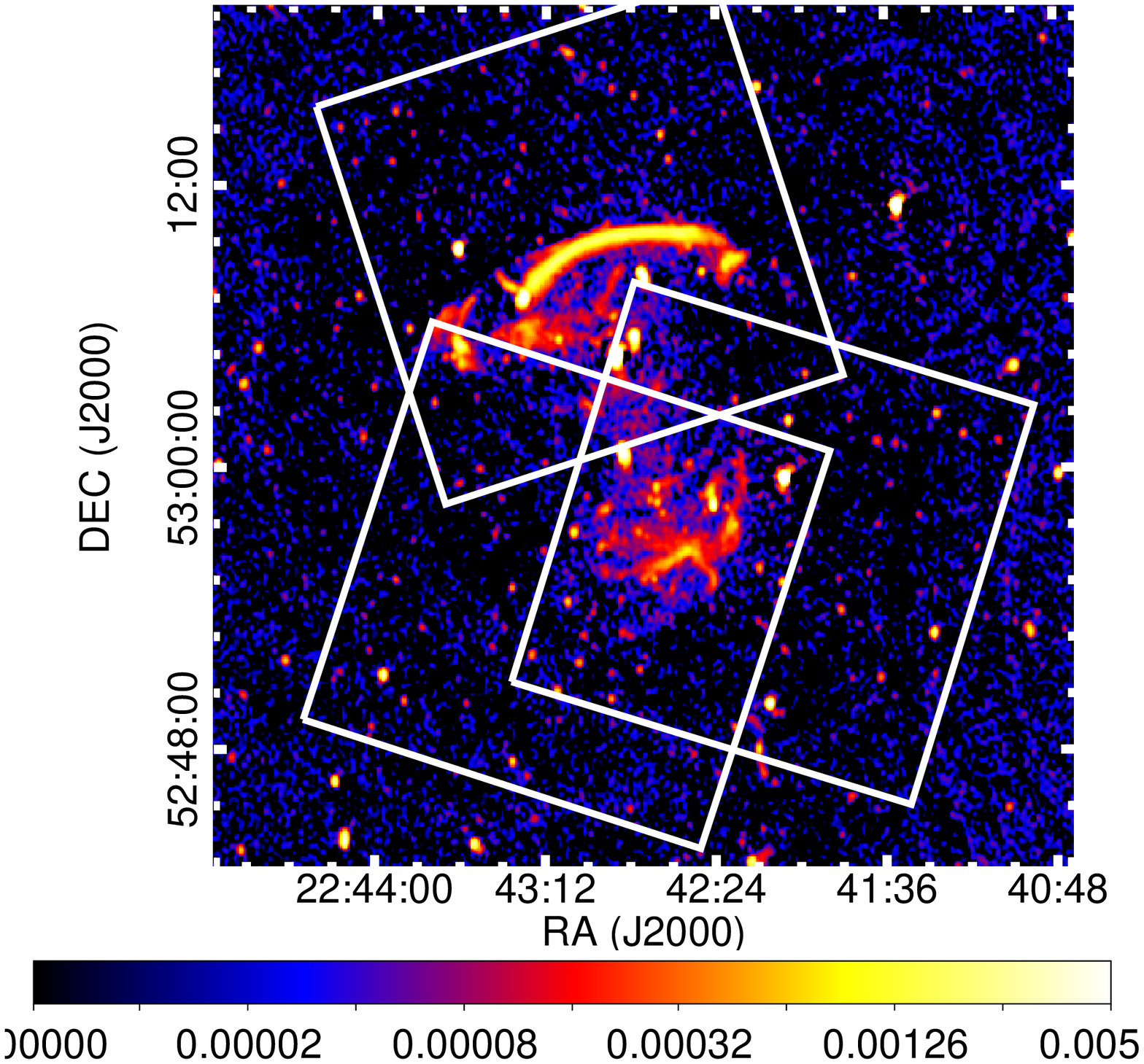}
\end{center}
\end{minipage}
\begin{minipage}{0.5\hsize}
\begin{center}
\includegraphics[width=1.\hsize]{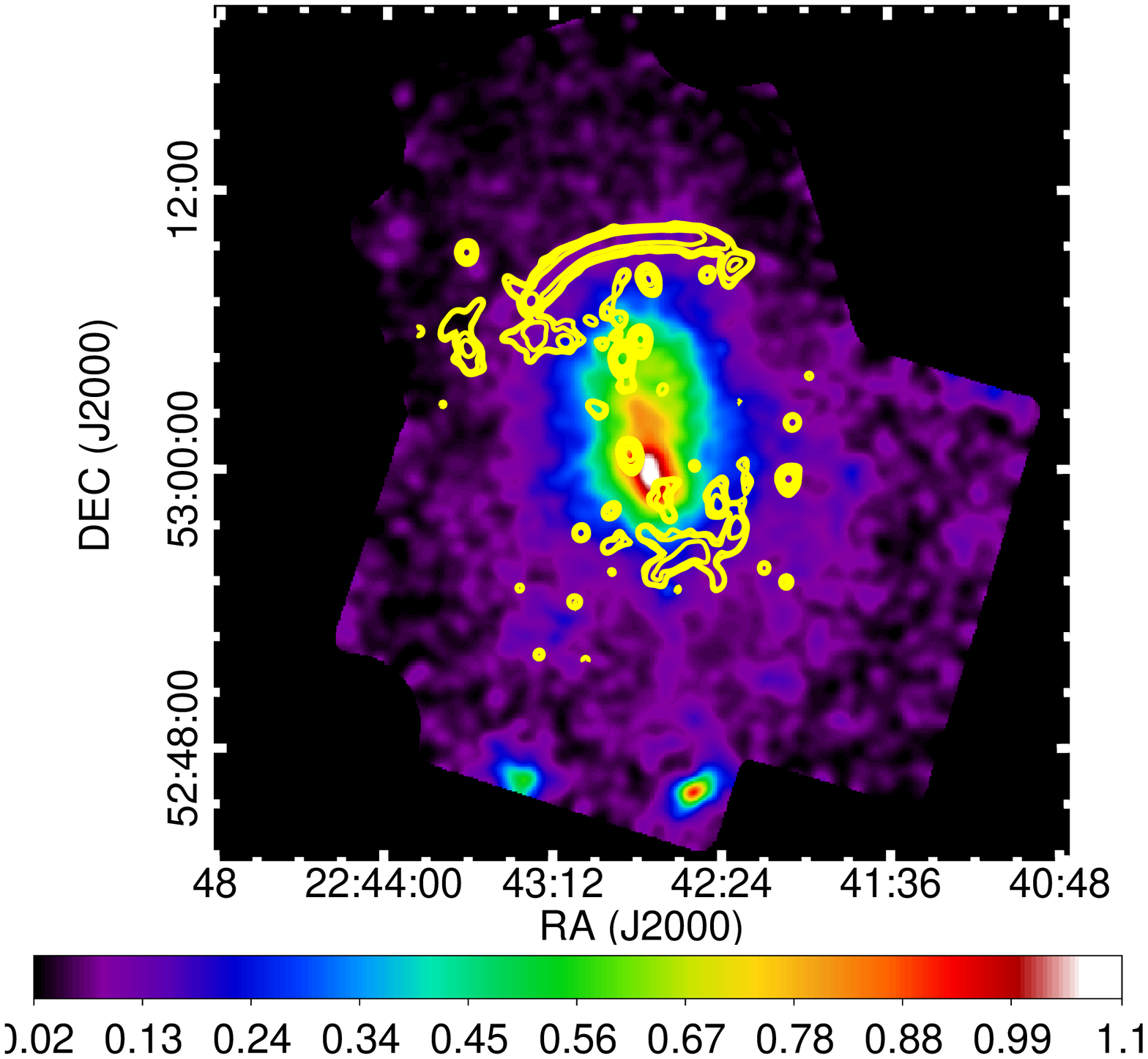}
\end{center}
\end{minipage}
\end{tabular}
\caption{Left:  {\it Suzaku} observations of CIZA J2242.8+5301 overlaid on a 1.4 GHz radio intensity map in color scale \citep{vanweeren10}. The {\it Suzaku} XIS fields of view are shown as white squares.
Right:
X-ray image of CIZA J2242.8+5301 in the energy band 0.5-2.0 keV, after
  subtraction of the NXB with no vignetting correction and after smoothing
  by a 2-dimensional gaussian with $\sigma$ =8 pixel =8.5\arcsec.   
  The 1.4 GHz radio emission is shown with yellow contours.  }
\label{fig:suzaku_image}
\end{figure*}

\section{Introduction}\label{sec:intro}
Based on the framework of the hierarchical structure formation theory, clusters of galaxies grow through merging events with smaller objects and accretion flows from large scale structure filaments~\citep{voit05, kravtsov12}. In particular, merging events between massive clusters of galaxies are the most energetic events in the Universe since the Big Bang, releasing as much as  $\sim$10$^{64}$ erg~\citep{markevitch07}.
 This gravitational energy is converted into heating of the Intra-Cluster Medium (ICM) and particle acceleration by shocks likely via the mechanism of diffusive shock acceleration \citep[DSA:][]{blandford87}.

Radio observations show the existence of diffuse non-thermal emission in clusters such as {\it radio~halos, relics} and {\it mini-halos} 
\citep[for a review, see][and references therein]{feretti12, brunetti14}.
 Among these radio structures, $radio~relics$ show remarkable features such as a diffuse (Mpc-scale), elongated shape, and they are typically located at cluster outskirts. In addition, radio relics are usually seen in merging clusters. Therefore, radio relics are considered to be the result of synchrotron emission from relativistic electrons which are accelerated by shocks induced by cluster mergers~\citep{ensslin98}.
 
Because shocks play a fundamental role in cluster evolution, it is crucial to understand their properties (Mach number, shock velocity, shock acceleration efficiency, etc.). 
In particular, at high Mach numbers (${\cal M}\sim10^3-10^4$) DSA is known to have enough efficiency to accelerate particles from a thermal distribution to the high energy regime. This is confirmed sby observations of several supernova remnants~
\citep[e.g.,][]{koyama95}. However, shock fronts induced by cluster mergers are expected to have a much lower Mach number (${\cal M}\sim2-5$). The acceleration efficiency of these low-Mach number shocks is thought to be too low to reproduce the observed radio brightness~\citep{kang12}.
However, recent work by~\cite{guo14a, guo14b}, using detailed particle in cell simulations indicates, that electron acceleration can be efficient at low-Mach number shocks in the ICM. More detailed studies of cluster shocks are highly desired for understanding their roles during merging activity.

Most evidence for shock fronts in clusters of galaxies is based on X-ray imaging studies \citep[for a review, see][and references therein]{markevitch10}. Up to now,  clear evidence for shock fronts are reported in 1E~0657-56 \citep[Bullet cluster:][]{markevitch02},  A520~\citep{markevitch05},  A754~\citep{macario11},  A2146~\citep{russell10,russell12}, A521~\citep{bourdin13} and A2034~\citep{owers14}.

X-ray follow-up observations of the northwest radio relic in Abell 3667 revealed that the ICM temperature and surface brightness distribution drops across the relic, which indicates the existence of a shock front across the relic~\citep{finoguenov10}. The correspondence between radio relics and shocks 
is also confirmed in other clusters~\citep{macario11, akamatsu12_a3667, akamatsu12_a3376,ogrean13_coma, bourdin13, akamatsu13b}. Using the Rankine-Hugoniot jump conditions, the estimated Mach number from the temperature jump typically spans ${\cal M}\sim1.5-3.0$~\citep{akamatsu13a}. These values are consistent with prediction from hydrodynamical simulations~\citep{miniati00, ryu03}.

This paper reports the results of deep X-ray observations of the cluster \object{CIZA J2242.8+5301} with ${\it Suzaku}$.  It has been discovered in the Clusters in the Zone of Avoidance (CIZA) survey 
\citep[$z=0.192$:][]{kocevski07}. \object{CIZA J2242.8+5301} has a clear double radio relic at a scale of several Mpc and with a  remarkably narrow width of 50 kpc for the northern relic~\citep{vanweeren10}. The relic is strongly polarized at the 50 to 60\% level, indicating a well-ordered magnetic field, which is well aligned with the shock plane.

The northern relic shows an injection index at the edge of the relic of -0.6$\pm0.05$~\citep{vanweeren10, stroe13}, which corresponds to a Mach number of ${\cal M}\sim 4.6$. 
The spectral index for the south relic is also reported as $-1.29\pm0.05$, which corresponds to a Mach number of ${\cal M}\sim 2.8$~\citep{stroe13}.
\citet{stroe14b} also report the detection of radio emission at the high frequency band ($\sim$16 GHz) and pose some questions for the diffusive shock acceleration predictions of the ageing effect on relativistic electrons.
A recent deep Chandra observation of \object{CIZA J2242.8+5301} revealed several edges in the X-ray surface brightness distribution. 
Interestingly, the edges are found in the downstream regions of the shocks 
assumed to be associated with the radio relics, and have no counterpart in the temperature distribution~\citep{ogrean_chandra}.

Previous ${\it Suzaku}$ observations of \object{CIZA J2242.8+5301} showed a remarkable temperature jump from 8.3 keV to 2.1 keV across the northern relic \citep{akamatsu13a}. The Mach number estimated from the Rankine-Hugoniot condition is ${\cal M}_{\rm X} = 3.2 \pm 0.5$. On the other hand, as described above, \citet{vanweeren10} reported a somewhat higher Mach number ${\cal M_{\rm Radio}}=4.6_{-0.9}^{+1.3}$. Similar higher values, ${\cal M}_{\rm radio}> {\cal M}_{\rm X}$, have been found for two other relics 
\citep[1RXS J0603:][Itahana in prep]{vanweeren12_toothbrush, ogrean13}.
 Recently, \citet{sarazin14} reported results of deep XMM-Newton observations of the northeast radio relic in Abell 3667. They derive a Mach number of 2.09$\pm$0.09, which is clearly much smaller than 
 the Mach number expected from the radio spectral index ~\citep[Spectrum index $\alpha$=-0.6$\pm$0.2, {\cal M}=4.6$\pm$0.3:][]{hindson14}. 

However, more recently \citet{stroe14c} reported the  possibility of underestimating the radio spectral index from spatially-resolved spectral fitting analysis for the northern radio relic in  \object{CIZA J2242.8+5301} and determine a new spectral index of $0.77^{+0.03}_{-0.02}$.
The Mach number estimated from the new radio spectral index is ${\cal M}=2.9_{-0.13}^{+0.10}$, which is now in agreement with X-ray observations~\citep{akamatsu13a, ogrean_chandra}.  

The combination of X-ray and radio observations gives us the opportunity to solve many problems related to shock physics, in particular the nature of the shock fronts themselves and their role in the acceleration and heating processes in the ICM. From this point of view, \object{CIZA J2242.8+5301} is an important target to study at X-ray wavelengths.

We use $H_0 = 70$ km s$^{-1}$ Mpc$^{-1}$, $\Omega_{\rm M}=0.27$ and
$\Omega_\Lambda = 0.73$, which gives a scale of 190 kpc per arcminute at $z = 0.192$.   We employ solar abundances defined by \citet{lodders03} and Galactic absorption with 
{N$_{\rm H:total}=46.0 \times10^{20}$~cm$^{-2}$~\citep{willingale13}}. 
Unless otherwise stated, the errors correspond to 68\% confidence for a single parameter.

\section{${\it Suzaku}$ observations \& Spectrum analysis}\label{sec:obs_data}
\begin{table}[]
\begin{center}
\caption{\label{tab:obslog}{\it Suzaku} observations log of the CIZA2242}
\begin{tabular}{cccccccccc}
\hline
Region & 	(R.A, DEC)& Exposure\\
                       &                     &(ks) \\
\hline
North & (340.74, 53.16)  & 82.2 \\  
East & (340.75, 52.92) & 74.0 \\
West & (340.51, 52.95) & 77.4 \\
OFFSET & (339.29, 52.67)  & 38.2   \\
\hline
\end{tabular}
\end{center}
\end{table}

\begin{table*}[t]
\caption{\label{tab:bbd-fit}Best-fit background parameters}
\begin{tabular}{lcccccccccccccccc} \hline
 & \multicolumn{2}{c}{Local Hot Bubble} &  \multicolumn{2}{c}{Milky Way Halo} & \multicolumn{2}{c}{Hot Foreground} & \multicolumn{2}{c}{CXB} \\  
 &  $kT$~(keV) & $norm^{\mathrm{a}}$ &  $kT$~(keV) & $norm^{\mathrm{a}}$ &  $kT$~(keV) & $norm^{\mathrm{a}}$ &  $\Gamma$& $norm^{\mathrm{b}}$ &  $\chi^2/d.o.f$ \\
 & &  $[\times10^{-3}]$ & & $[\times10^{-3}]$  & & $[\times10^{-3}]$  &  &$[\times10^{-4}]$ 
 \\ \hline
Case1&0.08 (fix) & $  {24.2}\pm2.2 $  & $  {0.30}^{+0.02}_{-0.01}  $  &  $  {4.9}\pm0.4  $ & -- & -- & 1.41 (fix) &  $  {9.4}\pm0.4  $  &  166/133 \\

{\bf Case 2} & 0.08 (fix) & $43.2\pm5.0$ & $0.27\pm_{-0.04}^{+0.01}$ & $6.1\pm1.5$ & $0.68_{-0.08}^{+0.04}$ &$1.0_{-0.3}^{+0.6}$ & 1.41 (fix) & $  {9.2}\pm0.4  $  & 159/132
\\
\hline
 \multicolumn{10}{l}{\footnotesize
$\mathrm{a}$: Normalization of the apec component scaled by a factor $20\times20\pi=400\pi$ (see text).}\\
\multicolumn{10}{l}{\footnotesize
$norm$=$\rm \frac{1}{400\pi}$$\int n_{e}n_{H} dV/(4\pi(1+z^2)D_{A}^2)\times 10 ^{-14} \rm ~cm^{-5}~arcmin^{-2}$, where 
$D_A$ is the angular diameter distance to the source.}\\
\multicolumn{10}{l}{\footnotesize
$\mathrm{b}$: The
CXB intensity normalization in Kushino et al. 2002 is  
9.6$~\times10^{-4}$ for $\Gamma=1.41$
in units of photons keV$^{-1}~\rm cm^{-2}~s^{-1}$ at 1 keV.} \\
\end{tabular}
\end{table*}

{\it Suzaku} carried out three observations of \object{CIZA J2242.8+5301} (Figure~\ref{fig:suzaku_image} left) and one observation of an OFFSET region, during the {\it Suzaku} AO-6 (PI: Kawahara) and AO-8 (PI:van Weeren) phase. We refer to these pointings as North, East and South, respectively. They cover the whole cluster beyond the radio relics. The observation log is given in Table~\ref{tab:obslog}.   All observations were performed with the normal 5$\times$5 and 3$\times$3 clocking  mode\footnote{http://www.astro.isas.ac.jp/suzaku/doc/suzaku\_td/node10.html} (without burst or windows options).

The XIS instrument consists of 4 CCD chips: one back-illuminated (BI: XIS1) and three front-illuminated  (FI: XIS0, XIS2, XIS3) chips~\citep{mitsuda07}. XIS2 was not operational due to damage from a meteoroid strike\footnote{http://www.astro.isas.ac.jp/suzaku/doc/suzakumemo/suzakumemo-2007-08.pdf}. Data reduction was done with {\tt HEASOFT} version 6.15 and CALDB version 20140203.

The XIS event lists created by the rev 2.5 pipeline processing were filtered using the following additional criteria: 
a geomagnetic cosmic-ray cut-off rigidity (COR) $> $ 8 GV,  and an elevation angle $\rm ELV >~12^\circ$.  We applied additional processing for the XIS1 detector to reduce the non-X-ray background (NXB) level, which has increased after the change in the amount of charge injection from 2 keV to 6 keV in June 2011. That operation improved the high-energy response with a negligible loss in the low-energy performance\footnote{http://www.astro.isas.ac.jp/suzaku/doc/suzakumemo/suzakumemo-2010-07v4.pdf}. The detailed processing procedures are the same as those described in $XIS~analysis~topics$\footnote{http://www.astro.isas.jaxa.jp/suzaku/analysis/xis/xis1\_ci\_6\_nxb/}. The 5$\times$5 and 3$\times$3 editing modes data formats were added. 

\begin{figure}[]
\begin{center}
 \includegraphics[scale=0.3,angle=-90]{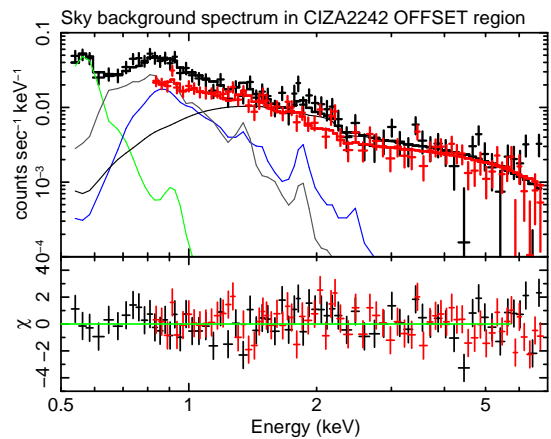}
\end{center}
\caption{\label{fig:bbd-spectrum}The spectrum of the OFFSET field used for the background estimation, after subtraction of the NXB and the point sources. The XIS BI (Black) and FI (Red) spectra are fitted with CXB + Galactic components (LHB, MWH and HF)   ({\it apec+ phabs(apec+powerlaw+apec)}).  The CXB spectrum is shown with a black curve, and the LHB, MWH and HF components
  are indicated by green, blue and gray curves, respectively.  }
\end{figure}

The observed spectrum was assumed to consist of optically-thin thermal plasma emission from the ICM, and emission due to the Galactic foreground, Cosmic X-ray background (CXB) and NXB\@.  
In order to investigate the properties of the cluster plasma outside of the radio relics, an accurate estimation of the foreground-background emission is essential. Here we use a {\it Suzaku} offset observation, which is located 1 degree east of \object{CIZA J2242.8+5301}, to estimate the sky background spectra. 

In the fitting procedure for the cluster emission, we combined the spectra from the FI detector (XIS0, 3)  and BI and FI spectra were fitted simultaneously. The NXB subtracted spectrum is fitted with a model of the ICM and the sky-background 
components. Details about background estimation and the ICM components across the northern/southern radio relics will be described in subsections~\ref{sec:bgd}, \ref{sec:north} and \ref{sec:south}. 

\begin{figure*}[th]
\begin{tabular}{cc}
\begin{minipage}{0.5\hsize}
\begin{center}
\includegraphics[width=1.\hsize]{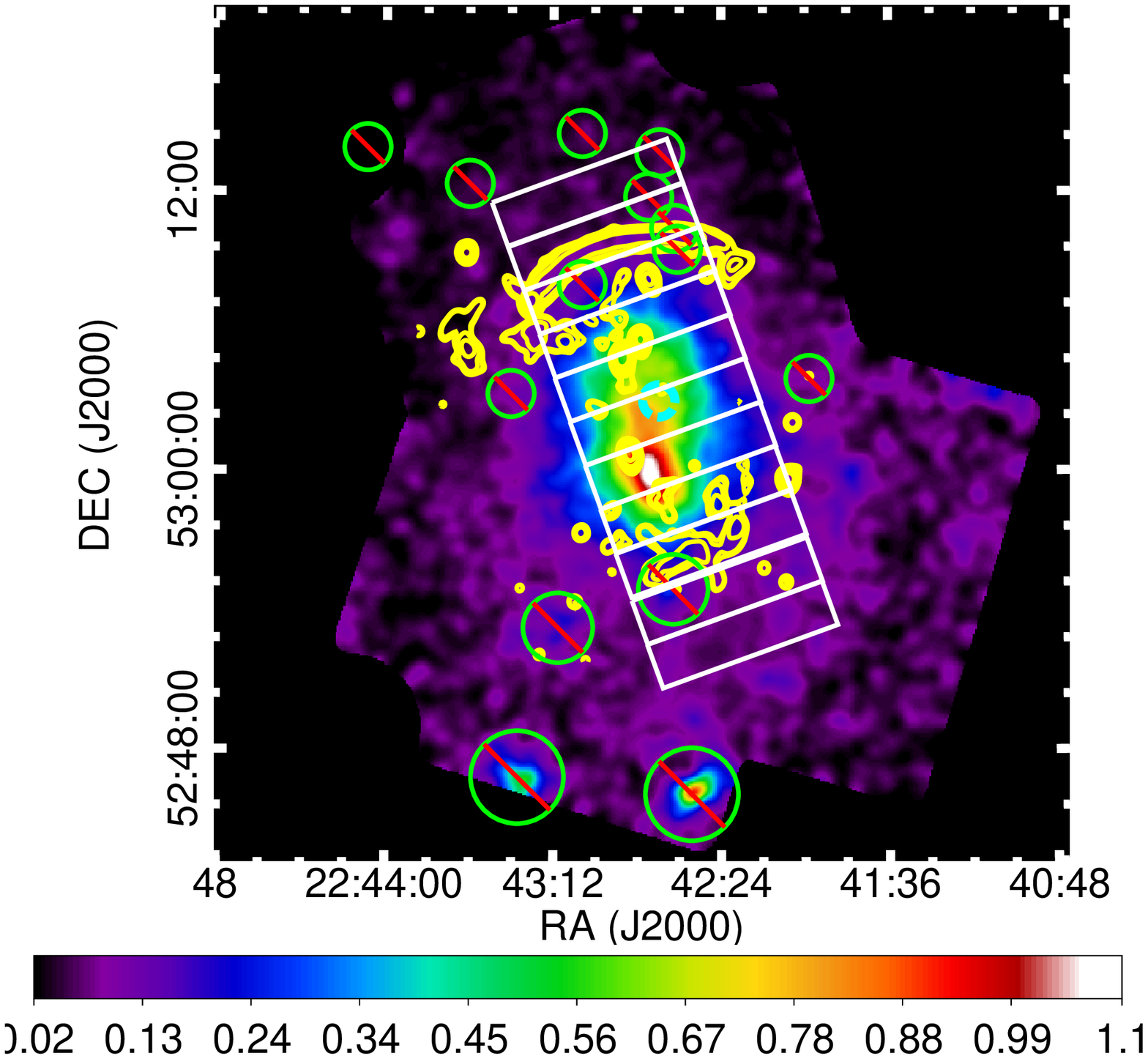}
\end{center}
\end{minipage}
\begin{minipage}{0.5\hsize}
\begin{center}
\includegraphics[width=.9\hsize,angle=-90]{Fig5.ps}
\end{center}
\end{minipage}
\end{tabular}
\caption{\label{fig:suzaku_image_box}
Left: X-ray image of the \object{CIZA J2242.8+5301} in the energy band 0.5-5.0 keV, after subtraction of the NXB with no vignetting correction and after smoothing by a 2-dimensional gaussian with $\sigma =8$ pixel =8.5\arcsec.  
The identified point sources by XMM-Newton are highlighted with the green circles (see text). The 1.4 GHz radio emission is shown with yellow contours. The dashed cyan circle indicate the center which was determined by visually fitting a circle to the radio relics.
Right: Temperature profile of \object{CIZA J2242.8+5301} from the  south to north
direction. The dotted gray lines show WSRT 1.4 GHz radio emission.
}
\end{figure*}

\subsection{Estimation of the background spectra}\label{sec:bgd}
The sky X-ray background is divided into four components: unabsorbed thermal emission from the Local Hot Bubble (LHB: $kT \sim$0.1 keV), absorbed thermal emission form the Milky Way Halo (MWH: $kT \sim$ 0.3 keV),  the Hot Foreground (HF: $kT\sim0.6$ keV) and  the CXB.

In order to determine the level of the sky X-ray background, we first analyzed  a {\it Suzaku} offset observation of \object{CIZA J2242.8+5301} (obsID:806002010). We assume that these sky background components are distributed uniformly over the {\it Suzaku} FOV. We used uniform ancillary response files (ARF) for the sky background components. The ARF for the sky background  was generated by using $xissimarfgen$~\citep{ishisaki07}  by assuming uniform-sky emission  over a circular region with a radius of 20\arcmin. Hereafter, we call this uniform-ARF.

The NXB component was reconstructed from the XIS Night Earth database with $xisnxbgen$ in {\tt FTOOLS}~\citep{tawa08}. We accumulated the data for the same detector area and the same distribution of COR2 as the observations. To increase the signal to noise ratio, we applied data selection with COR2 $>$ 8 GV~
\citep[see Figure 3 in ][]{tawa08}. To adjust for the long-term variation of the XIS background due to radiation damage, we select the night Earth data within 300 days before and after the period of the observation.
Using the $wavdetect$ function in the CIAO package\footnote{http://cxc.harvard.edu/ciao/ahelp/wavdetect.html}, we searched for point sources in the {\it Suzaku} image. 
Point sources were detected down to $\sim1\times10^{-13}$~erg~cm$^{-2}$~s$^{-1}$ in the 2-10 keV band. We exclude the point sources within a 1-2 arcmin radius in order to take account for the point spread function (PSF) of the {\it Suzaku} XRT~\citep{xrt}. 

For the spectral fits, we used the XSPEC version~12.8.0 package\footnote{http://heasarc.nasa.gov/xanadu/xspec/}. The model for the sky-background components is described as  $Apec_{\rm LHB}+phabs(Apec_{\rm MWH}+Apec_{\rm HF}+powerlaw_{\rm CXB})$ in XSPEC representation.  In this model, {\it Apec} and {\it phabs} represents thin thermal plasma emission model~\citep{apec} and Galactic absorption toward the target, respectively.  We fixed the abundance and the redshift of each thermal component to 1.0 and 0.0, respectively. Further we fixed the temperature of the LHB and photon index of the CXB components to 0.08 keV and  $\Gamma$=1.41~\citep{kushino02}, respectively. We used the spectra in the 0.5-7~keV range for the BI detectors and 0.8-7~keV for the FI detector. The result is consistent with the CXB intensity with the ~\citet{kushino02} level and the MWH temperature of {0.27$_{-0.04}^{+0.01}$} keV is consistent with typical values in other fields~\citep{yoshino09}. 
A fit with the HF component fixed to zero  (case1)  is sightly poorer than a fit with its temperature and normalisation as free parameters (case2). The resulting fit parameters
are shown in Table ~\ref{tab:bbd-fit}. Basically these  models reproduce all features of the observed spectrum well,
CXB intensity, low energy ($<$ 1~keV)  sky-background.
We have checked  how the HF component improves the fitting with an F-test, and obtained the probability value of the F-test $\sim$ {0.001}. Thus, we employed the parameters of case 2 in Table~\ref{tab:bbd-fit}.  The best-fit spectrum model (case2) is shown in Figure~\ref{fig:bbd-spectrum}. 

\begin{figure*}[t]
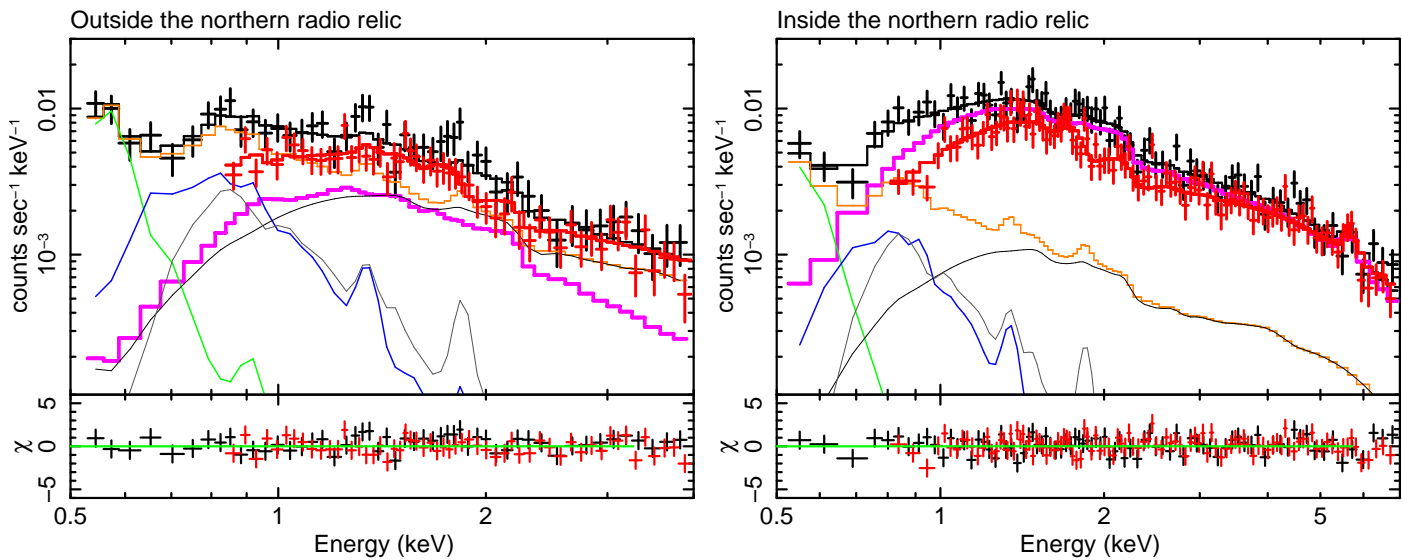

\begin{tabular}{cc}
\begin{minipage}{0.5\hsize}
\centering
 \includegraphics[width=.8\hsize,angle=-90]{Fig6.ps}
 \end{minipage}
\begin{minipage}{0.5\hsize}
\centering
 \includegraphics[width=.8\hsize,angle=-90]{Fig7.ps}
 \end{minipage}
\end{tabular}
\caption{ \label{fig:fit}
NXB subtracted spectra in each annular region. The XIS BI (Black) and FI (Red) spectra are fitted with the ICM model ({\it phabs $\times$ apec}), along with the sum of the CXB and the Galactic emission 
of Case 2 ({\it apec +  phabs(apec + apec + powerlaw)}). 
The ICM component is shown with a magenta line.
The CXB component is shown with a black curve, and the LHB, MWH and HF emissions are indicated by green, blue and gray curves, respectively. The sum of all sky background components is shown by the orange curve.
The Left panel shows the spectrum of the outside region. Right panel shows the inside region.
}
\end{figure*}


\begin{figure*}[th]
\begin{tabular}{cc}
\begin{minipage}{0.5\hsize}
\begin{center}
\includegraphics[width=1.\hsize]{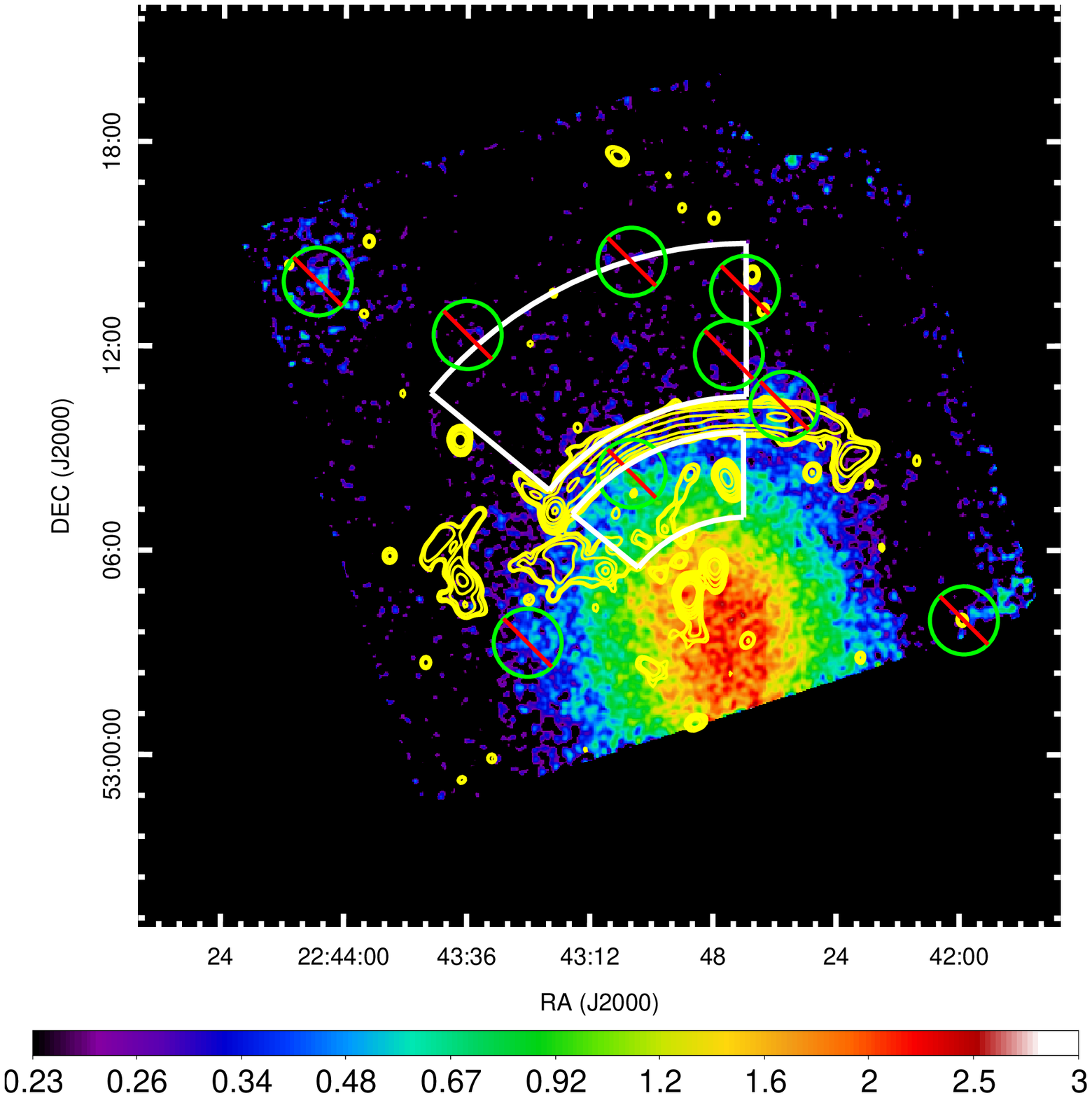}
\end{center}
\end{minipage}
\begin{minipage}{0.5\hsize}
\begin{center}
\includegraphics[width=1.\hsize,angle=-90]{Fig9.ps}
\end{center}
\end{minipage}
\end{tabular}
\caption{\label{fig:suzaku_north}
Left: {\it Suzaku} XIS image for the northern part of \object{CIZA J2242.8+5301}.  
Right: Radial profile of the ICM temperature for the north direction. The present best-fit values with statistical errors are shown with black crosses.  Gray dotted crosses indicate previous {\it Suzaku} results, which do not include HF component~\citep{akamatsu13a}.  The range of uncertainties due to the combined 3\% variation of the NXB level and 
 the maximum/minimum fluctuation in the CXB is shown by two green dashed lines.The black dotted lines show the WSRT 1.4 GHz radio intensity scaled arbitrary in flux. The higher peak corresponds the northern radio relic and the second peak is a point source just behind the radio relic.
}
\end{figure*}

\subsection{ICM emission along the merging direction}\label{sec:box}
In order to investigate the cluster temperature distribution in the merging direction (from north to south), we divided the cluster into 11 box regions as shown in Figure~\ref{fig:suzaku_image_box}.
For the ICM emission around the radio relic,  we employed an absorbed single temperature thermal plasma model ($phabs\times apec$). In all regions, we fixed the metal abundance and the redshift to 0.3~\citep{fujita08,werner13} and 0.192~\citep{kocevski07}, respectively. Because the ICM emission has a different spatial distribution from the sky-background, we generated an ARF by using the {\it Suzaku} XIS image (0.5-8.0 keV band) as input surface distribution for $xissimarfgen$. Based on Sec.~\ref{sec:bgd}, we modeled the sky-background  emission using the uniform ARF. Employing multiple ARFs is enabled with XSPEC version 12. The normalization of the LHB component, the normalization and temperature of the  MWH component, and the normalization of the power-law model for the CXB component were allowed to vary within the range of the errors in the background estimation~(table~\ref{tab:bbd-fit}). We carried out spectral fits to the pulse-height spectra in each annular region separately.  In the simultaneous fit of the BI and FI data, only the normalizations were allowed to be different between them, 
although we found that
the derived normalizations were consistent within 15\%
 The resulting temperature profile is shown in Figure~\ref{fig:suzaku_image_box}. The temperature profile shows a significant drop at  $r$=-6 and 6 arcmin, respectively. The temperature drop strongly indicates the presence of shock fronts at those locations. Here we note that  the temperature profile was smeared by the point spread function of the {\it Suzaku} XRT~\citep[half power diameter~$\sim$~2 arcmin:][]{xrt}.
Therefore, we can not resolve structures of scales less than an arcmin. In the next section, we will investigate in detail the properties around the radio relics.

\subsection{ICM emission across the northern radio relic}\label{sec:north}
To investigate the detailed ICM properties associated with the radio relics we extracted pulse-height spectra in two annular regions whose boundary radii were 4.0\arcmin-7.0\arcmin~and 7.5\arcmin-11.5\arcmin~for the  region inside and outside the radio relic with the center at (22:42:41.9, 53:03:00.0), shown in Figure~\ref{fig:suzaku_north}. To reduce contamination from the brighter region,  we take the 
region outside the radio relic 0.5\arcmin~away from the outer edge of the inner region.
To maximize the signal to noise ratio outside the northern relic region, we used the spectra in the 0.5--4~keV range for the 
BI detector and 0.8--4~keV for the FI detectors.

To investigate a possible systematic error in the temperature estimation, we consider a 3\% fluctuation of the NXB level and fluctuations of the CXB intensity as shown below. It is important to exclude the contribution of point sources in the region of interest, to reduce the systematic error of the CXB  caused by fluctuations of unresolved sources. For the elimination of the point sources, we used the archival data of XMM-Newton (OBSID=0654030101). As shown by small (green) circles in Figure~\ref{fig:suzaku_image_box}, we detected several point sources and subtracted 10 point sources with 2.0-10.0 keV fluxes higher than $S_c=1\times10^{-14}\rm~erg~cm^{-1}~s^{-1}$. According to previous studies of the CXB with the ASCA satellite~\citep{kushino02}, the CXB flux fluctuation in  the field of view of the ASCA GIS (0.5 deg$^2$) is 6.5\% in the 2-10 keV band. We scaled the measured fluctuations from our flux limit and the observed area following \citet{hoshino10}. The estimated fluctuations are 32\% and 18\% for inside and outside of the radio relic regions, respectively. 
We repeated all the spectral fits by fixing the CXB intensity at the upper and lower boundary values.

The pulse-height spectra and the best-fit models for both annular regions are shown in Figure~\ref{fig:fit}.  We obtained fairly good fits for all the regions with reduced $\chi^2$ values less than 1.2. The resultant ICM temperature with different background models are listed in table~\ref{tab:mach}. The temperature changes significantly from $kT_{\rm inside}={8.5}^{+0.8}_{-0.6}$ keV to $ kT_{\rm outside}={2.7}_{-0.4}^{+0.7}$ keV. Even though we use different sky background models,  these results are consistent with previous {\it Suzaku} results \citep[$kT_{\rm inside}={8.3}\pm0.8$ keV and $kT_{\rm outside}={2.1}\pm0.4$ keV:][]{akamatsu13a}. 
These ICM temperatures are consistent with the ones estimated from the box regions shown in Figure~\ref{fig:suzaku_image_box}.

For the inside and outside the northern relic, box region shows ${kT_{\rm inside}= 9.7\pm1.3}$ keV and ${kT_{\rm outside}=3.4\pm1.1}$ keV, respectively. The ICM temperature of the post-shock region also agrees with a previous XMM-Newton observation 
 \citep[$kT_{\rm inside}\sim8-11$ keV:][]{ogrean13}.
Figure~\ref{fig:suzaku_north} shows the radial profile of the ICM temperature across the northern radio relic. Black crosses show the best fit ICM temperature of each annular region. Grey dotted crosses represent previous results~\citep{akamatsu13a}, which do not include the HF component. 

One can also estimate the shock compression from the jump in the X-ray surface brightness across the relic.
The surface brightness was derived from the observed spectra. 
We checked the surface brightness across the well defined northern radio relic.
The resulting un-absorbed X-ray surface brightness (0.5-2.0 keV) shows a factor 8 difference across the relic, which corresponds to a shock compression of $C\sim \sqrt{8.22}=2.86$, including the difference in emissivity for the two regions. 
This value matches the compression parameter ($C\sim\sqrt{7.41} = 2.72$) derived across the rectangular sector, which is shown Figure~\ref{fig:suzaku_image_box}.
These surface brightness jumps are consistent with the shock compression predicted based on the temperature jump  (Table.5: $C\sim2.83_{-0.20}^{+0.34}$). This agrees with the results for shocks found on other clusters (for instance, Bullet: Markevitch et al. 2002, Shimwell et al. 2015, A520: Markevitch et al. 2005, A754:Macario et al. 2011, RXJ1314.4-2515 Mazzotta  et al. 2011).
However, we note that our estimation based on our X-ray spectroscopic analysis represent the upper limit to the density jumps because the surface brightness profiles of clusters of galaxies typically have steep power-law profiles. It is difficult to make concrete statements on this issue due to the limited spatial resolution of the {\it Suzaku} XRT.

\begin{table}[]
\small
\caption{\label{tab:pre}Best-fit ICM temperature across the northern radio relic with different background model (unit of keV)}
\begin{center}
\begin{tabular}{lcccccccccccccccc} \hline
& Case1	& {\bf Case2$^{\mathrm{a}}$} &Case3{$^\mathrm{b}$} \\ \hline
Outside N relic & 2.7$_{-0.4}^{+0.5}$	&
2.7$_{-0.4}^{+0.7}$$_{-0.3}^{+0.5}$ &
2.6$_{-0.5}^{+0.9}$\\
Inside N relic & 8.3$_{-0.5}^{-0.7}$ &
8.5$_{-0.6}^{+0.8}$$_{-0.2}^{+0.2}$&
8.6$\pm0.9$
\\
\hline
\multicolumn{4}{l}{\footnotesize 
$\mathrm{a}$: The first error is the statistical one and the second one
}\\
\multicolumn{4}{l}{\footnotesize 
is due to systematics one.
}\\
\multicolumn{4}{l}{\footnotesize 
$\mathrm{b}$: Case 2 background model with $>$ 1 keV energy band.
} \\
\end{tabular}
\label{tab:north_kT}
\end{center}
\end{table}
\begin{table}[]
\small
\caption{\label{tab:pre}Best-fit ICM temperature across the southern radio relic with different background model (unit of keV)}
\begin{center}
\begin{tabular}{lcccccccccccccccc} \hline
& Case1	& {\bf Case2$^{\mathrm{a}}$} &Case3{$^\mathrm{b}$}\\ \hline
Beyond S relic & 
4.4$_{-0.3}^{+0.8}$ &  
5.0$_{-0.6}^{+0.9}$$_{-0.9}^{+1.1}$ & 
5.2$_{-0.9}^{+2.0}$
\\
Inside S relic & 
8.8$_{-0.3}^{+0.4}$ &  
9.0$\pm$0.5$\pm$0.4 & 
9.0$\pm0.6$
\\
\hline
\multicolumn{4}{l}{\footnotesize 
$\mathrm{a}$: Idem as above.
}\\
\multicolumn{4}{l}{\footnotesize 
$\mathrm{b}$: Idem as above.
}\\
\end{tabular}
\label{tab:south_kT}
\end{center}
\end{table}

\begin{figure*}[]
\begin{tabular}{cc}
\begin{minipage}{0.5\hsize}
\centering
 \includegraphics[scale=0.45]{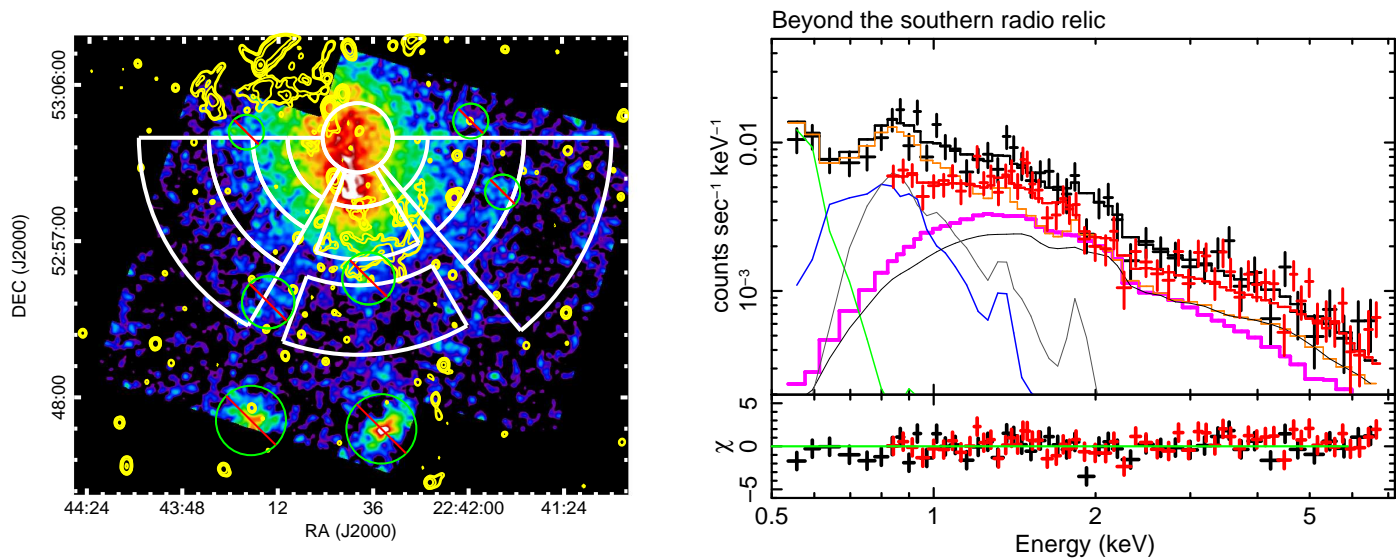}
  \end{minipage}
\begin{minipage}{0.5\hsize}
  \includegraphics[width=.8\hsize,angle=-90]{Fig11.ps}
   \end{minipage}
   \end{tabular}
\caption{ \label{fig:spec_south}
Radial profile of the ICM temperature for the southern direction. Colors are the same as in Figure.~\ref{fig:suzaku_north}.}
\end{figure*}
\begin{figure*}[]
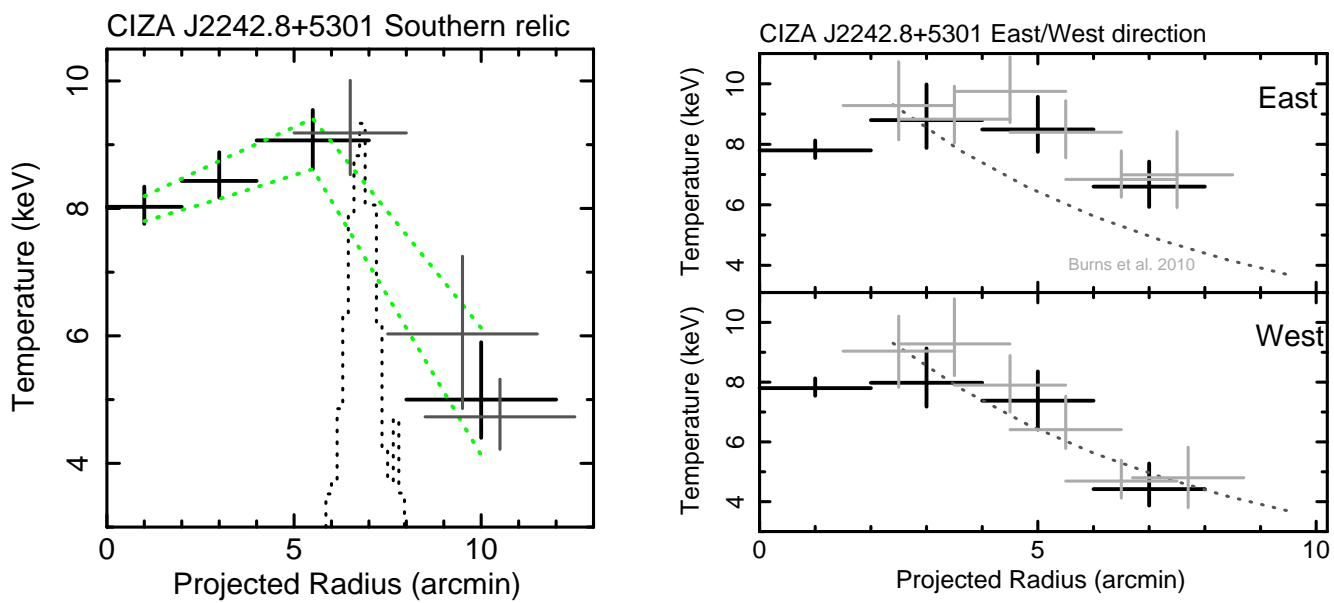

\begin{tabular}{cc}
\begin{minipage}{0.5\hsize}
\centering
 \includegraphics[scale=0.4,angle=-90]{Fig12.ps}
  \end{minipage}
\begin{minipage}{0.3333\hsize}
 \includegraphics[scale=0.4,angle=-90]{Fig13.ps}
   \end{minipage}
   \end{tabular}
\caption{ \label{fig:profile_south}
Radial profile of the ICM temperature for the South (left), East and West direction. {The Black crosses show the best-fitted values and the gray crosses show the the results of slightly offset regions. }
The gray dotted curves in the right panels show the temperature profile \citep{burns10} (see text).  In the left panel, we removed the radio point sousrce located just behind the relic. 
}
\end{figure*}

\subsection{ICM emission around the southern radio relic}\label{sec:south}
Next we investigate the ICM properties around the southern radio relic
 with our new {\it Suzaku} observations. To increase the signal with our limited observation time, the West and East observations overlap at the southern radio relic. 
We analyzed these data in the same way as the northern one.  
Similar as for the northern relic, we take a one arcmin gap to avoid contamination from the brighter part of the cluster due to the PSF of the {\it Suzaku} XRT. 
To investigate possible structures in the W and E directions, we also evaluate the ICM temperatures by shifting the annuli by 0.5\arcmin. The resulting ICM temperatures are shown by the gray crosses in the Figure 7.

We used the spectra in the 0.5--7~keV range for the BI detectors and 0.8--7~keV for the FI detector. For point source exclusion, we only used {\it Suzaku} data  because XMM-Newton/Chandra observations did not completely cover the full {\it Suzaku} FOV. We excluded point sources by using CIAO to a flux detection limit of $S_c=1\times10^{-13}$~erg~cm$^{-2}$~s$^{-1}$. Based on the flux detection limit, we estimate the possible fluctuation of the CXB intensity in each region. The estimated fluctuations span 35-60~\%. 

We successfully obtained the ICM temperature across  and beyond the south relic for the first time. The resulting ICM temperature across the radio relic region is shown  in Table~\ref{tab:south_kT}.  The best-fit temperature profiles are shown in Figure~\ref{fig:profile_south}. Toward the southern radio relic, the ICM temperature show slightly increasing profile from {8.0$\pm$0.3} keV to {9.0$\pm$0.5} keV and shows a significant drop to {5.0$_{-0.6}^{+0.9}$} keV. 
Similar to the results for the northern shocks/relic, these temperatures are consistent with those derived for in the rectangular sector (Figure~\ref{fig:suzaku_image_box}), although the statistical errors are large.
This might be due to the fact that this region is located too far from the radio relic, the bin is placed about 760 kpc from the peak of radio emission. Given the expected decrease in temperature with radius this would explain the lower temperature measured there.
Combined with the fluctuation of the NXB level \citep[3~\%:][]{tawa08}, we investigated the systematic error. The results are shown by green dotted lines in Figure~\ref{fig:profile_south}.  Due to the high point source flux detection limit, the fluctuations of the CXB intensity are large. The emission intensity outside of the southern radio relic is almost comparable to the sky background (Figure~\ref{fig:spec_south}). Therefore, the systematic errors outside of the southern radio relic are larger than the statistical error.

\section{Discussion}\label{sec:discussion}
We performed deep {\it Suzaku} observations of the double radio relic cluster \object{CIZA J2242.8+5301}. We measure the ICM temperature profiles beyond the relics were measured. The profiles show a significant drop across the radio relics, which implies the presence of shock fronts.  We discuss the full cluster temperature structure by comparing to other clusters. We evaluate the shock properties (Mach number, compression factor and propagation speed), the possibility of non-equilibrium effects due to shock heating and we compare these results with the radio observations.

\begin{table*}[t]
\caption{Shock Properties at the northern and southern radio relic}
\centering
\begin{tabular}{ccccccccc}\hline \hline
&$T_2^\mathrm{a}$ & $T_1^\mathrm{a}$ & Mach No.$^\mathrm{b}$ & $v_{shock}^\mathrm{c}$ & Compression & {Power-law} slope$^\mathrm{d}$ & Spectrum index\\
&(keV) & (keV) & ${\cal M}$ & (km s$^{-1}$)  & $C$ 	&  $p$ & $\alpha$\\ \hline
Northern &
{$8.5_{-0.6}^{+0.8}$} & {$ 2.7_{-0.5}^{+0.9}$}& {$ 2.7_{-0.4}^{+0.7}$} &
{$ 2300^{+700}_{-400}$} & {$ 2.84_{-0.24}^{+0.43}$} & 
{$ 2.64\pm 0.30$} &{ $0.81\pm0.15$}  \\

Southern &{$9.0\pm0.6$} &

{$5.1_{-1.2}^{+1.5}$} & {$1.7_{-0.3}^{+0.4}$} & {$2040^{+550}_{-410}$} & {$2.01_{-0.34}^{+0.45}$}&{$3.96_{-0.99}^{+1.32}$} & {$1.48_{-0.49}^{+0.66}$}
\\ \hline
\multicolumn{8}{l}{1,2 indicate pre-shock and post-shock, respectively.}\\
\multicolumn{8}{l}{$\mathrm{a}$: 
{
The error ranges calculated by considering  statistical and systematic errors (see text in Sect.3.2 for details).
}
}\\
\multicolumn{8}{l}{$\mathrm{b}$: From the Rankine-Hugoniot temperature jump condition
$\frac{T_2}{T_1} = \frac{5{\cal M}^4+14{\cal M}^2-3}{16{\cal M}^2} $}\\
\multicolumn{8}{l}{$\mathrm{c}$: The shock speed $v_{shock}={\cal M}\cdot v_{s}$, 
$\mathrm{d}$: Expected Power-law slope $p=(C+2)/(C-1)$}
\end{tabular}
\label{tab:mach}
\end{table*}

\begin{figure}[]
\begin{center}
    \includegraphics[width=.55\hsize, angle=-90]{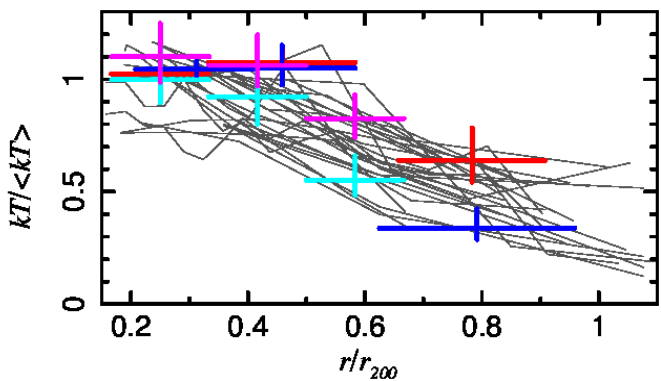}
\end{center}
\caption{Scaled projected temperature profiles compared with relaxed clusters~\citep[Gray: see Fig.~8 in][]{reiprich13}. The profiles were normalized to the average temperature. The $r_{200}$ value was derived from~\citet{henry09}.
Blue, red, yellow and magenta crosses show the temperature profile of the North, South, East and West direction,  respectively.
}
\label{fig:avg_temp}
\end{figure}

\subsection{Large scale ICM temperature structure}
As suggested by numerical simulations, clusters grow via merging with sub-clusters. These merging activities have a huge impact on the temperature structure~\citep[see for a review ][]{markevitch07}. Recent {\it Suzaku} observations reveal the ICM temperature profiles up to the virial radius~
\citep[see for a review ][]{reiprich13}. However, these studies mainly focus on relaxed clusters and  studies of merging cluster outskirts are still limited except for a few examples 
\citep[e.g.,][]{ibaraki14}. Therefore, the impact of merging activity on the cluster structure remains poorly known. 

The new {\it Suzaku} observations of \object{CIZA J2242.8+5301}, in combination with {\it Suzaku} observations taken in 2011~\citep{akamatsu13a}, cover the full cluster region.  
As shown in Figure~\ref{fig:suzaku_image_box},  \object{CIZA J2242.8+5301} has a peculiar temperature profile along the merging axis (from north to south). 
Contrary to relaxed clusters~\citep{reiprich13}, \object{CIZA J2242.8+5301} has an almost flat radial profile with $kT\sim~8$ keV \citep[][and this work]{ogrean13}.  In addition, across the radio relics, the temperature profile also shows a significant drop.  These observed temperature profiles strongly indicate the presence of a shock front and shock heating on the ICM. We discuss the properties of the shock fronts in the next subsection.

We plotted a model temperature profile proposed by \citet{burns10} as gray dotted curves in the right panel of Figure.~\ref{fig:profile_south}. Their model profile reproduces well the measured ICM temperature structure of relaxed clusters of galaxies by the {\it Suzaku} satellite \citep{reiprich13}. For the calculation of the model profile, we adopt $T_{\rm avg}=8.0 $ keV and $r_{200}=12$ arcmin, respectively. Here, $r_{200}$\footnote{$r_{200}$ is the radius where the mean total density of the cluster is 200 times the mass density of the Universe  with critical density.} is estimated from~\citet{henry09}.
We also compare the measured temperature profiles with previously measured clusters~(Figure~\ref{fig:avg_temp}). The temperature profile in the West direction agrees with the model profile. On the other hand, the profile in the East direction shows an excess.  The temperature excess in the East direction is consistent with the previous XMM-Newton results~\citep{ogrean13}. They report the presence of a high temperature ($kT\sim$13 keV) region at the eastern  part of the cluster. In our temperature profile, we do not find such high a temperature region. However, due to the limited PSF of the {\it Suzaku} XRT, we may miss such a component. In addition,  such a high temperature ($kT> 10$ keV) is beyond the effective energy band of XIS and of XMM-Newton.
Therefore tight constraints on high temperature components are difficult to obtain for current X-ray CCD detectors. NuSTAR\footnote{http://www.nustar.caltech.edu/}~\citep{harrison10} and the upcoming ASTRO-H\footnote{http://astro-h.isas.jaxa.jp/index.html.en}~\citep{takahashi12}  satellite will help to detect such high temperature components with their hard X-ray imaging and spectral capabilities.

\subsection{Shock properties}
The {\it Suzaku} temperature profiles of \object{CIZA J2242.8+5301} show clear drops across the relics, indicating the presence of shock fronts. We evaluate the shock properties at both radio relics based on the {\it Suzaku} results. 
Hereafter we used the error including systematic errors 
due to the variation of the NXB level and 
 the maximum/minimum fluctuation in the CXB
defined as $\sigma_{\rm err}\equiv\sqrt{\sigma_{Stat}^2+\sigma_{\rm Sys}^2}.
$
We cannot determine a one-to-one 
relationship between the edge of the radio relics and the shock fronts because of the limited PSF of the {\it Suzaku} XRT. 

The Mach number can be obtained by applying the Rankine-Hugoniot jump condition
\citep{landau59}

\begin{eqnarray}
\frac{T_2}{T_1} = \frac{5{\cal M}^4+14{\cal M}^2-3}{16{\cal M}^2}, 
\end{eqnarray}
assuming the ratio of specific heats as $\gamma=5/3$.   Here, 1,2 indicate pre-shock and post-shock, respectively. Table~\ref{tab:mach} shows the resultant Mach numbers based on the temperature jump, whose values are $ {\cal M_{\rm n}}=2.7_{-0.4}^{+0.7}$, ${\cal M_{\rm s}}=1.7_{-0.3}^{+0.4}$, respectively (Table~\ref{tab:mach}).

\label{sec:shock}
 \begin{figure}[]
\begin{center}
    \includegraphics[width=1.\hsize]{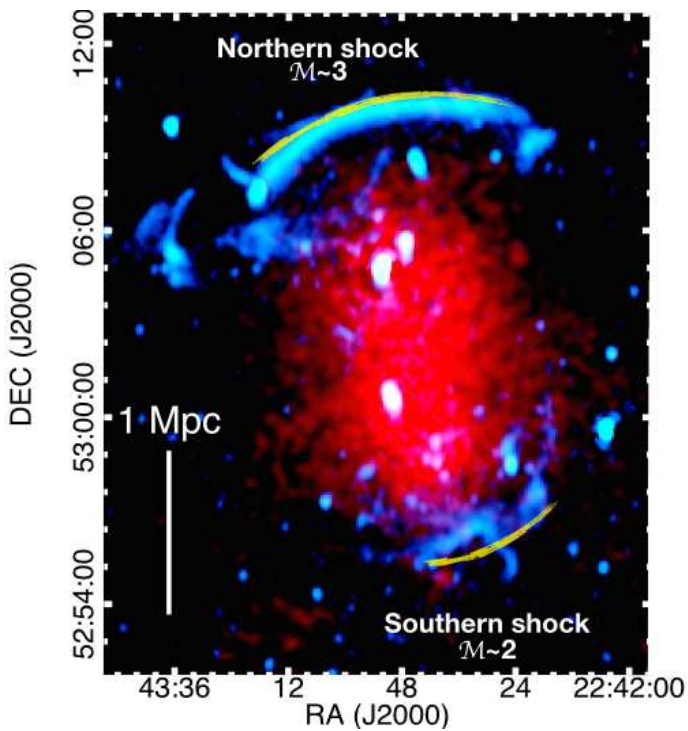}
\end{center}
\caption{ Smoothed 0.5-2.0 keV band X-ray image (red) and WSRT 1.4 GHz image of \object{CIZA J2242.8+5301} (cyan). 
The thin yellow lines depict the approximate locations of the shock fronts confirmed by ${\it Suzaku}$.
\label{fig:suzaku_threecolor_image}
}
\end{figure}

Using the measured temperature in the pre shock region ($kT_{\rm pre:n}\sim2.7$ keV, $kT_{\rm pre:s}\sim5.1$ keV),  the sound speed is $c_{\rm s: n}\sim 850\rm~ km~ s^{-1}$, $c_{\rm s: s}\sim 1170\rm~ km~ s^{-1}$.   With these Mach numbers, the expected shock compression $C$ and  the shock propagation speed  ($v_{\rm shock}={\cal M}\cdot c_{\rm s}$) are estimated  as 
{$C_{\rm n}=2.84_{-0.24}^{+0.43}$}, 
{$C_{\rm s}=2.01_{-0.34}^{+0.45}$} and  
{$v_{shock:\rm n}=2300_{-400}^{+700}~\rm km~s^{-1}$}, 
{$v_{shock:\rm s}=2040_{-410}^{+550}~\rm km~s^{-1}$}, respectively. For the northern radio relic, the estimated shock speed is in agreement with the radio observations~\citep[$v_{shock:n}\sim2500\rm~km~s^{-1}$:][]{stroe14c}.
Here we estimate the shock compression $C$  from the Mach numbers derived from the temperature drops.
Although the Mach number is comparable to that of the Bullet cluster (${\cal M}\sim3.0$), the shock velocity in \object{CIZA J2242.8+5301} is much smaller than for the Bullet cluster ($4500$$\rm~km~s^{-1}$) and is comparable to A520 (2300$\rm~km~s^{-1}$) and other systems (e.g. A2034: 2057$\rm~km~s^{-1}$).
The relative gas flow speed to the shock front is $v_{gas:n}=v_{shock}/C\sim810~\rm km~s^{-1}$ and $v_{gas:s}\sim1000~\rm km~s^{-1}$.  The estimated downstream velocity for the northern shock is in agreement with the one used for estimating the magnetic field \citep[$v_{gas}=1000~\rm km~s^{-1}$, $v_{gas}=905_{-125}^{+165}~\rm km~s^{-1}$:][]{vanweeren10, stroe14c}.

Here we discuss possible systematics on our estimate of pre shock temperature.
The Mach number we determined is in principle also affected by any temperature gradient that was present before the shock passage. Any shock heating will appear as an excess above the temperature gradient that was already present. We can provide an approximate estimate of this effect by assuming the gradient before shock passage followed the profile from~\citet{burns10}. Substituting an average temperature of <{\it kT}>=8 keV into Eq.~8 of \citet{burns10}, the expected temperature gradient between the center of pre and post shock is ~2.0 keV. So the temperature at the current post-shock location, but before shock passage was about 2.7 + 2.0 = 4.7 keV. After passage of the shock this increased to the currently measured post-shock value of 8.5 keV. With a value of T$_1$=4.7 keV and T$_2$=8.5 keV this corresponds to a Mach number of 1.8.

Additional uncertainties arise from the fact that we assumed the temperature measurements are valid for the center of the bins in the above calculation. Our actual temperature estimate is emission weighted, resulting in a temperature bias in the high density region. Therefore, the position of the weighted temperature of the pre-shock bin could actually be located closer to the relic position (see fig.~6 in \cite{hoshino10}) and the shock heating is expected to be located at the edge of the relic. Due to this fact, the correction for the pre-shock temperature is likely somewhat smaller than what we assumed above. As an example, the expected temperature gradient between the edge of the relic and the center of the pre-shock region is $\sim$0.8 keV, assuming the same~\citet{burns10} profile.  The expected pre-shock temperature precisely at the edge of the shock would then be 2.7+0.8=3.5 keV. With our measured post-shock temperature this would then corresponds to a Mach number of 2.3. We stress that the above calculations are just estimates as we lack information on the original temperature profile before the merger. The discussion in the following sections and conclusions are not affected by this systematics.

\subsection{Merger scenario of \object{CIZA J2242.8+5301}}\label{sec:merger_sysnario}
The temperature distribution of the ICM shown in Figure~\ref{fig:suzaku_image_box}, \ref{fig:suzaku_north}, \ref{fig:profile_south} and the presence of the shock fronts suggest that a subcluster is colliding into the main body from the North-South direction. The peak of X-ray emission is close to the southern part of the cluster~\citep{ogrean13, ogrean_chandra}, which might be infalling from the north direction. In this point of view, the northern shock corresponds to the backward shock induced by the merging activity.
We do not see a significant excess in the temperature profile 
in the perpendicular direction to the merger axis, 
which is predicted by several simulations~\citep{akahori10, molnar13}.
Overall, our observations indicate that \object{CIZA J2242.8+5301} is experiencing an almost head-on merger with near unity mass ratio and small impact parameter.

Here we estimate the dynamical time scale of \object{CIZA J2242.8+5301} by using its shock properties and radio information.
Assuming that the shock propagation speed does not change from $v_{shock}\sim2000$ km/s and that the distance from the north to the south relic is 2.5 Mpc, we estimate that  \object{CIZA J2242.8+5301} is observed approximately 2.5 Mpc/(2$\times v_{shock})\sim$0.6 Gyr after core passage. Note that the actual times scale is expected to be longer as the shock speed just after core passage is lower.
Therefore in the end, the total travel time would be longer than 0.6 Gyr.
Our results support previous estimates based on hydrodynamical simulation and radio observations~\citep[][]{vanweeren11_sim, stroe14c}.
They suggested that \object{CIZA J2242.8+5301} is a binary cluster merger with a mass ratio of 1:2,  less than 10 degrees from the plane of the sky
and impact parameter $b < 400$ kpc. The combination of radio and X-ray observations is powerful to constrain cluster merger events.  For a more detailed understanding of the dynamical evolution of the merger,  the mass distribution from a weak lensing observation and hydrodynamical simulations are useful \citep[e.g.,][]{dawson13}.

\subsection{Mach number from X-ray and Radio observations}\label{sec:comp}
Recent X-ray follow-up observations of the radio relics revealed a relation between the shock front and radio emission~\citep{finoguenov10, macario11, mazzotta11, ogrean13,ogrean13_coma,  akamatsu13a}. These results are consistent with the hydrodynamical simulations of merging clusters, which predict that merging events generate shock waves toward outskirts~\citep{takizawa98,ricker01,kang05}. 
These shocks accelerate electrons up to relativistic energies via diffusive shock acceleration (DSA) mechanism, which generates radio emission via synchrotron radiation.  However,  as mentioned in the introduction,  the acceleration  efficiency of DSA  with low-M ($<10$) shocks is known to be too low  to take account for the radio flux.
Therefore the main acceleration mechanism at cluster merger shocks is still unclear and until now the subject of discussion. For understanding of shock acceleration, it is crucial to know the particle acceleration efficiency and injection rate. 

\begin{figure}[]
\begin{center}
    \includegraphics[width=1.\hsize]{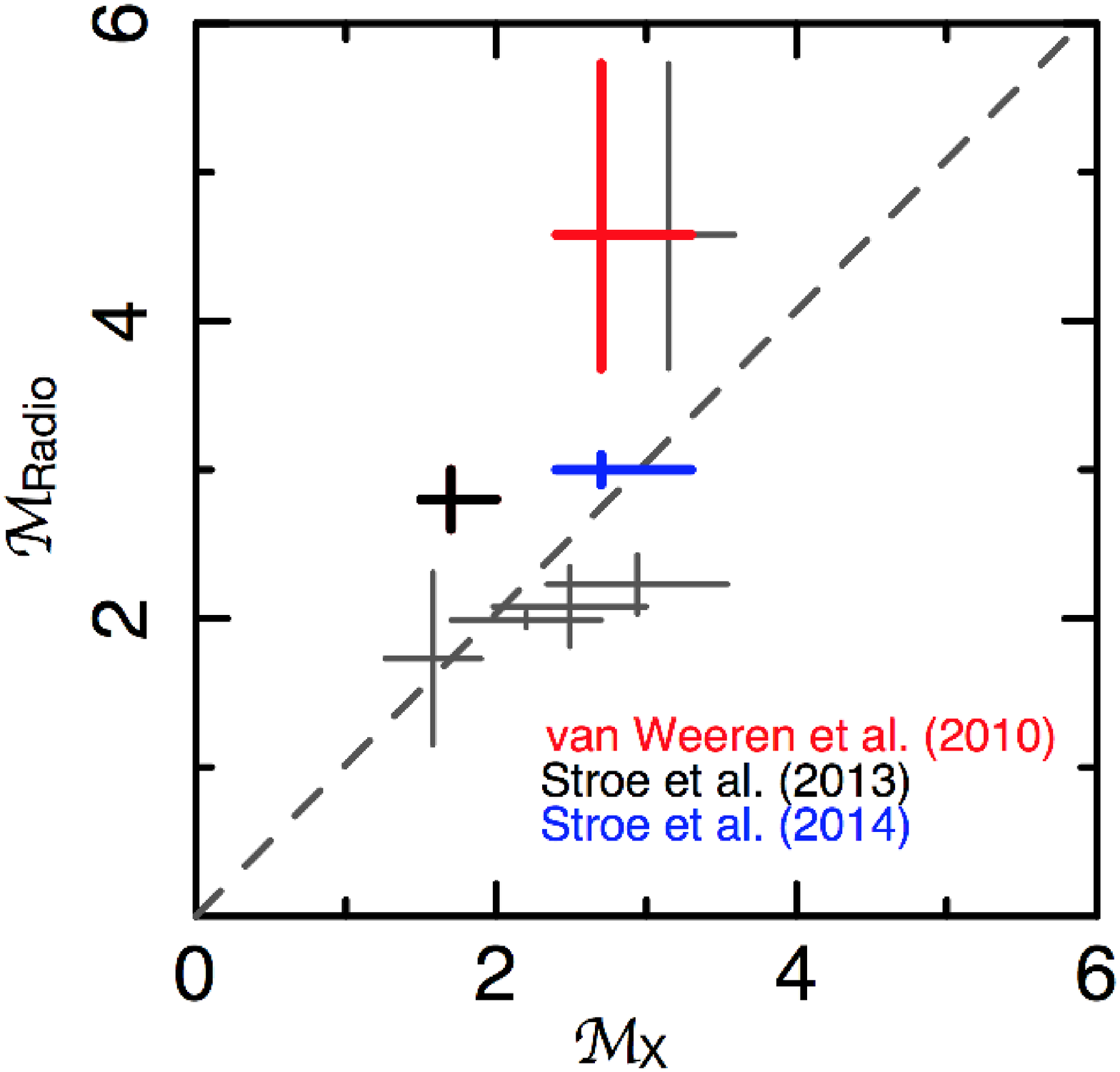}
\end{center}
\caption{The Mach number derived from radio (Mradio) plotted against that from the ICM temperature. 
The error bars are only statistical and at the one sigma level.
Red and blue crosses are this work, but with previous and new radio result~\citep{vanweeren10, stroe14b}. 
Grey crosses show the previous our results~\citep{akamatsu13a}.
Note: \citet{stroe14c} does not give information about the southern relic because the low flux. Therefore the result of the new radio analysis is only available for the northern relic.
However, similar to the northern relic case, the Mach number of the southern radio relic is expected to drop.
\label{fig:mach}
}
\end{figure}

Because of the presence of the two clear and giant radio relics, \object{CIZA J2242.8+5301} is one of  the best targets to study shock acceleration in clusters. Especially, the northern relic shows an extremely narrow shape, whose width is only 50 kpc (Figure~\ref{fig:suzaku_image}). \citet{vanweeren10,  stroe13} reported the spectral index map of the radio relics in \object{CIZA J2242.8+5301}. At the edge of the northern radio relic, the spectral index is estimated from the spectral index map and color-color diagram as $\alpha_n=-0.6\pm0.05$, which correspond to the Mach number of ${\cal M}=4.6^{+1.3}_{-0.9}$. They also confirm the presence of a spectral index gradient across the relic, which is consistent with the DSA  theory combined with Synchrotron aging.
On the other hand, for the southern radio relic, it was not possible to estimate the injection spectral index directly from spectral index maps due to poor signal-to-noise.
Therefore, the estimated injection spectral index from the measured integrated spectral index $\alpha_{int}=-1.29\pm0.05$ as $\alpha_{inj}=-0.79\pm0.05$, which corresponds to a Mach number of ${\cal M}=2.8\pm0.2$~\citep{stroe13}.
We note that the estimated injection spectral index remains uncertain due to the assumptions that have to be made if it is derived from the integrated spectrum~\citep{kang15b, kang15a}.
The new results of the radio analysis~\citep{stroe14c} is slightly different from the previous result~\citep[$\alpha_n\sim-0.6$:][]{vanweeren10, stroe13}. The main difference is that \cite{stroe14c} used a largest radio data set and took into account spatial convolution effects. Because of the low flux of the southern radio relic, \cite{stroe14c} could not estimate the spectral index. Therefore the result from the new radio analysis is only available for the northern relic.

First order Fermi acceleration (under the assumptions of stationary and continuous injection) gives relativistic electrons with a power-law spectrum $n(E)dE\sim E^{-p}dE$, where the power-law index is $p=(C+2)/(C-1)$.  The radio spectral slope just after the shock acceleration is $\alpha=-(p-1)/2$. Using the measured shock compression, this relation implies 
{$p_n=2.64\pm0.30$}, 
{$p_s=3.96_{-0.99}^{+1.32}$} and 
{$\alpha_n=-0.81\pm0.15$}, 
{$\alpha_s=-1.48_{-0.49}^{+0.66}$}, respectively. 
For the northern radio relic, the above value is consistent with  the spectral index from the radio investigation~\citep[][$\alpha\sim-0.6\pm0.05$, $\alpha\sim-0.77$]{vanweeren10, stroe14c}.
For the DSA theory, the measured injection index would require a shock compression of $C_n\sim2.84$, which is again consistent with what is suggested by the temperature drop across the radio relic (Table.~\ref{tab:mach}: $C_n\sim2.8$).

Although it is likely to change similar to the northern radio relic case, there is a difference between the X-ray and radio measurement of the southern radio relic. Using the measured integrated spectral index and an assumption for the cooling effect,  a similar trend has been seen in several radio relics and edges of the halo~\citep{markevitch05, ogrean13_coma, ogrean13, akamatsu13b}. There are several mechanism proposed to explain the particle acceleration in clusters~\citep{markevitch05, kang12, pinzke13, skillman13}. A plausible possibility is re-acceleration of pre-existing non-thermal (low energy cosmic-ray) particles in the ICM. In this case, the pre-shock ICM already contains several type of non-thermal particles, which are generated by several mechanisms such as large scale accretion shock~\citep{ryu03}, turbulence acceleration~\citep{brunetti11} and past activity of radio or starburst galaxies  \citep[see][for a review]{bruggen12}. These low energy cosmic-ray particles have a long life time comparable to the ages of clusters or longer~\citep{sarazin99, pinzke13}.  Recently, ~\citet{kang12} performed  time-dependent simulations for diffusive shock acceleration for the case of the northern radio relic in \object{CIZA J2242.8+5301}. Compared with the observed synchrotron flux and spectral distributions, they concluded that if a pre-existing population of low-energy cosmic-ray electrons exists,  the radio relics can be explain by weak shocks with a Mach number ${\cal M}\sim2-3$. A similar conclusion is also reported by \citet{pinzke13}. They found that  contributions of fossil electrons can not be ignored for low Mach number shocks and their self-similar analytic model described the observed radio properties well.  Another possibility is that of a varying Mach number along the shock. The  ICM in the cluster outskirts is thought to be clumpy~\citep{nagai11, simionescu11}, the inhomogeneities in the ICM will thus result in small Mach number variations which could lead to a discrepancy between the radio and X-ray derived Mach numbers as the shock acceleration efficiency (and radio relic brightness) scales non-linearly with Mach number~\citep{hoeft07}.

For \object{CIZA J2242.8+5301}, we measured the ICM temperature jump across the northern relic, which corresponds to a Mach number of ${\cal M}\sim 3$. 
Indeed, the latest radio investigations by \citet{stroe14c} suggest that the previous radio  spectral index might be biased and that the estimated spectral index will be $\alpha\sim-0.77$, which correspond to a Mach number ${\cal M}\sim3.0$.
The new Mach number from radio observation agrees well with X-ray observations.  
In addition, the measured Mach number matches the prediction of the simulation by \citet{kang12}. This supports  the scenario of re-acceleration of the pre-existing non-thermal particles in the ICM. Upcoming LOFAR low-frequency radio observations will shed new light on this problem. Such a high quality radio spectrum allows us to observe the injection spectral index directly. 

\subsection{Magnetic field pressure at the northern radio relic}\label{sec:pressure}
As suggested by previous radio observations~\citep{vanweeren10}, the northern relic has a relatively high magnetic field ($B\sim 5\rm~\mu G$). Similar results are also reported for other radio relics~\citep{nakazawa09}. Such a magnetic field strength is comparable to the expected value at the cluster central region from simulations of Faraday Rotation Measure in the Coma cluster~\citep{bonafede10}. 

The thermal pressure of the ICM and the magnetic pressure of relativistic electrons is calculated  as $P_{\rm th}=kT\cdot n_e$ and $P_{magnetic}=B^{2}/(8\pi)$, respectively.
Assuming the ICM temperature and electron density $kT\sim8.0$ keV and $n_e\sim 5\times10^{-4}$~cm$^{-3}$, 
{we estimate the thermal pressure at the northern radio relic $P_{th}\sim$6.7 eV/cm$^3$, assuming that the mean molecular weight is 0.6. Here we assume that all X-ray emission comes from the thermal component. With a magnetic field strength of B $\sim$ 5 $\mu$G, the energy density is estimated as $P_{magnetic}\sim0.62$ eV/cm$^3$. Comparing both pressures, we find that the magnetic pressure reaches $\sim 9$ \% of the thermal pressure. In the above estimate, we did not include the contribution of the energy density of the non-thermal electrons and turbulence. Therefore $P_{B}/P_{th}\sim0.09$  is a robust lower limit of the non-thermal pressure.
}

\subsection{Possibility of non-equilibrium effects}\label{sec:nei}
Finally we discuss the possibility of non-equilibrium effects.
Similar to supernova remnants, cluster merger shocks are expected to be collisionless.  In the shock in young SNR, there is a well known relation between  the ratio of the ion and electron temperature $T_e/T_i$ and shock velocity as  $\displaystyle T_e/T_i\sim v_{shock}^{-2}$~\citep{vanadelsberg08}. In the cluster case,  non-equilibrium effects such as non-Maxwellian electron distributions~\citep{kaastra09} and  non-equipartition of electrons and ions temperature ~\citep{takizawa05, rudd09, akahori10} and non-equilibrium ionization are also expected. 

Because of the different mass between the ions and the electrons, in general, shock heating is more effective for the ions. The electrons set the same temperature via Coulomb collisions. After that, electrons stabilize down to a Maxwellian distribution. Finally, via ion to ion collisions, the ionization balance in the ICM reaches equilibrium.
The collisional time scale is given by $n_et_{ie}=3\times10^{11}~\rm cm^{-3}~s$.
Here, $n_et_{ie}$ is the ionization parameter, which is an indicator of the ionization degree~\citep{masai84}.
The region just behind the shock
may have a higher ion temperature $T_i$ than the electron one
$T_e$, because it has not had enough time to reach equilibrium.
We also tried to evaluate the ionization parameter $n_et$ at the post-shock region using the NIE model in Xspec.
Unfortunately {\it Suzaku} can not constraint well the $n_et$ parameter because of the  low abundance and low statistics of Fe-K lines.
Further studies should be possible by high-resolution spectroscopy of the ICM
with the Japanese ASTRO-H~\citep{takahashi12} and the European ATHENA\footnote{http://www.the-athena-x-ray-observatory.eu/}~\citep{nandra13} satellites. 

\section{Summary}
\object{CIZA J2242.8+5301} ($z=0.192$) hosts two well defined giant radio relics. We find that the temperature profiles show a remarkable drop across both relics. Below is a summary of our main results:

\begin{itemize}
\item The ICM temperature of the merging direction shows a flat profile with $kT\sim 8$ keV and shows a clear drops to 3-5 keV across the radio relics. The ICM temperature of the perpendicular direction to the merging axis shows different profiles. 
The temperature profile in the western direction agrees with the simulated temperature profile of \cite{burns10}.
\item The significant drop in the ICM temperature indicates the presence of shock fronts. Based on the temperature drop, the estimated Mach number of the shocks are 
{${\cal M}_n=2.7^{+0.7}_{-0.4}$}, and 
{${\cal M}_s=1.7_{-0.3}^{+0.4}$}, respectively. The shock velocities are also estimated as $v_{shock:n}\sim2300$ km/s and $v_{shock:s}\sim2000$ km/s, respectively. This suggests that the merger happened $\sim$0.6 Gyr ago, in agreement with estimates based on hydrodynamical simulation and dynamical analysis~(\cite{vanweeren11_sim, stroe14c} and Dawson et al. in prep.).
\item The observed radio spectral index of the northern relic~\citep{stroe14c} is consistent with  what would be expected from diffusive shock acceleration in a shock with the observed Mach number and compression.
\item By combining our X-ray data with radio results, we estimated the fraction of the magnetic pressure to the thermal pressure at the northern radio relic to be about $\sim0.62$ eV/cm$^{3}$, which corresponds to  9\% of the thermal pressure.
\end{itemize}

\bigskip The authors wish to thank the referee for constructive comments that significantly improved the manuscript.
The authors thank the {\it Suzaku} team members for their support of the {\it Suzaku} project and T. H. Reiprich for the data to reproduce temperature profiles obtained with {\it Suzaku}. We would also like to thank A. Simionescu and K. Sato for useful discussions.

H.A. is supported by a Grant-in-Aid for Japan Society for the Promotion of Science (JSPS) Fellows (26-606). R.J.W. is supported by NASA through the Einstein Postdoctoral grant number PF2-130104 awarded by the Chandra X-ray Center, which is operated by the Smithsonian Astrophysical Observatory for NASA under contract NAS8-03060.
HK is supported by the Astrobiology Project of the CNSI, NINS (AB261006) and by Grant-in-Aid for Scientific research from JSPS and from the MEXT (No. 25800106).
AS acknowledges financial support from NWO.
DS acknowledges financial support from the Netherlands Organisation for Scientific research (NWO) through a Veni fellowship, from FCT through a FCT Investigator Starting Grant and Start-up Grant (IF/01154/2012/CP0189/CT0010) and from FCT grant PEst-OE/FIS/UI2751/2014. 
MH acknowledges financial support by the DFG, in the framework of the DFG Forschergruppe 1254 `Magnetisation of Interstellar and Intergalactic Media: The Prospects of Low-Frequency Radio Observations'.
G.A.O. acknowledges support by NASA though a Hubble Fellowship grant
HST-HF2-51345.001-A awarded by the Space Telescope Science Institute,
which is operated by the Association of Universities for Research in
Astronomy, Inc., under NASA contract NAS5-26555.
SRON is supported financially by NWO, the Netherlands Organization for Scientific Research.

\bibliographystyle{aa}
\bibliography{ciza_aa}

\begin{thebibliography}{82}
\expandafter\ifx\csname natexlab\endcsname\relax\def\natexlab#1{#1}\fi

\bibitem[{{Akahori} \& {Yoshikawa}(2010)}]{akahori10}
{Akahori}, T. \& {Yoshikawa}, K. 2010, \pasj, 62, 335

\bibitem[{{Akamatsu} {et~al.}(2012{\natexlab{a}}){Akamatsu}, {de Plaa},
  {Kaastra}, {Ishisaki}, {Ohashi}, {Kawaharada}, \&
  {Nakazawa}}]{akamatsu12_a3667}
{Akamatsu}, H., {de Plaa}, J., {Kaastra}, J., {et~al.} 2012{\natexlab{a}},
  \pasj, 64, 49

\bibitem[{{Akamatsu} {et~al.}(2013){Akamatsu}, {Inoue}, {Sato}, {Matsusita},
  {Ishisaki}, \& {Sarazin}}]{akamatsu13b}
{Akamatsu}, H., {Inoue}, S., {Sato}, T., {et~al.} 2013, \pasj, 65, 89

\bibitem[{{Akamatsu} \& {Kawahara}(2013)}]{akamatsu13a}
{Akamatsu}, H. \& {Kawahara}, H. 2013, \pasj, 65, 16

\bibitem[{{Akamatsu} {et~al.}(2012{\natexlab{b}}){Akamatsu}, {Takizawa},
  {Nakazawa}, {Fukazawa}, {Ishisaki}, \& {Ohashi}}]{akamatsu12_a3376}
{Akamatsu}, H., {Takizawa}, M., {Nakazawa}, K., {et~al.} 2012{\natexlab{b}},
  \pasj, 64, 67

\bibitem[{{Blandford} \& {Eichler}(1987)}]{blandford87}
{Blandford}, R. \& {Eichler}, D. 1987, \physrep, 154, 1

\bibitem[{{Bonafede} {et~al.}(2010){Bonafede}, {Feretti}, {Murgia}, {Govoni},
  {Giovannini}, {Dallacasa}, {Dolag}, \& {Taylor}}]{bonafede10}
{Bonafede}, A., {Feretti}, L., {Murgia}, M., {et~al.} 2010, \aap, 513, A30

\bibitem[{{Bourdin} {et~al.}(2013){Bourdin}, {Mazzotta}, {Markevitch},
  {Giacintucci}, \& {Brunetti}}]{bourdin13}
{Bourdin}, H., {Mazzotta}, P., {Markevitch}, M., {Giacintucci}, S., \&
  {Brunetti}, G. 2013, \apj, 764, 82

\bibitem[{{Br{\"u}ggen} {et~al.}(2012){Br{\"u}ggen}, {Bykov}, {Ryu}, \&
  {R{\"o}ttgering}}]{bruggen12}
{Br{\"u}ggen}, M., {Bykov}, A., {Ryu}, D., \& {R{\"o}ttgering}, H. 2012, \ssr,
  166, 187

\bibitem[{{Brunetti}(2011)}]{brunetti11}
{Brunetti}, G. 2011, \memsai, 82, 515

\bibitem[{{Brunetti} \& {Jones}(2014)}]{brunetti14}
{Brunetti}, G. \& {Jones}, T.~W. 2014, International Journal of Modern Physics
  D, 23, 30007

\bibitem[{{Burns} {et~al.}(2010){Burns}, {Skillman}, \& {O'Shea}}]{burns10}
{Burns}, J.~O., {Skillman}, S.~W., \& {O'Shea}, B.~W. 2010, \apj, 721, 1105

\bibitem[{{Dawson}(2013)}]{dawson13}
{Dawson}, W.~A. 2013, \apj, 772, 131

\bibitem[{{Ensslin} {et~al.}(1998){Ensslin}, {Biermann}, {Klein}, \&
  {Kohle}}]{ensslin98}
{Ensslin}, T.~A., {Biermann}, P.~L., {Klein}, U., \& {Kohle}, S. 1998, \aap,
  332, 395

\bibitem[{{Feretti} {et~al.}(2012){Feretti}, {Giovannini}, {Govoni}, \&
  {Murgia}}]{feretti12}
{Feretti}, L., {Giovannini}, G., {Govoni}, F., \& {Murgia}, M. 2012, \aapr, 20,
  54

\bibitem[{{Finoguenov} {et~al.}(2010){Finoguenov}, {Sarazin}, {Nakazawa},
  {Wik}, \& {Clarke}}]{finoguenov10}
{Finoguenov}, A., {Sarazin}, C.~L., {Nakazawa}, K., {Wik}, D.~R., \& {Clarke},
  T.~E. 2010, \apj, 715, 1143

\bibitem[{{Fujita} {et~al.}(2008){Fujita}, {Tawa}, {Hayashida}, {Takizawa},
  {Matsumoto}, {Okabe}, \& {Reiprich}}]{fujita08}
{Fujita}, Y., {Tawa}, N., {Hayashida}, K., {et~al.} 2008, \pasj, 60, 343

\bibitem[{{Guo} {et~al.}(2014{\natexlab{a}}){Guo}, {Sironi}, \&
  {Narayan}}]{guo14a}
{Guo}, X., {Sironi}, L., \& {Narayan}, R. 2014{\natexlab{a}}, \apj, 794, 153

\bibitem[{{Guo} {et~al.}(2014{\natexlab{b}}){Guo}, {Sironi}, \&
  {Narayan}}]{guo14b}
{Guo}, X., {Sironi}, L., \& {Narayan}, R. 2014{\natexlab{b}}, \apj, 797, 47

\bibitem[{{Harrison} {et~al.}(2010){Harrison}, {Boggs}, {Christensen}, {Craig},
  {Hailey}, {Stern}, {Zhang}, {Angelini}, {An}, {Bhalerao}, {Brejnholt},
  {Cominsky}, {Cook}, {Doll}, {Giommi}, {Grefenstette}, {Hornstrup}, {Kaspi},
  {Kim}, {Kitaguchi}, {Koglin}, {Liebe}, {Madejski}, {Kruse Madsen}, {Mao},
  {Meier}, {Miyasaka}, {Mori}, {Perri}, {Pivovaroff}, {Puccetti}, {Rana}, \&
  {Zoglauer}}]{harrison10}
{Harrison}, F.~A., {Boggs}, S., {Christensen}, F., {et~al.} 2010, in Society of
  Photo-Optical Instrumentation Engineers (SPIE) Conference Series, Vol. 7732,
  Society of Photo-Optical Instrumentation Engineers (SPIE) Conference Series

\bibitem[{{Henry} {et~al.}(2009){Henry}, {Evrard}, {Hoekstra}, {Babul}, \&
  {Mahdavi}}]{henry09}
{Henry}, J.~P., {Evrard}, A.~E., {Hoekstra}, H., {Babul}, A., \& {Mahdavi}, A.
  2009, \apj, 691, 1307

\bibitem[{{Hindson} {et~al.}(2014){Hindson}, {Johnston-Hollitt},
  {Hurley-Walker}, {Buckley}, {Morgan}, {Carretti}, {Dwarakanath}, {Bell},
  {Bernardi}, {Bhat}, {Bowman}, {Briggs}, {Cappallo}, {Corey}, {Deshpande},
  {Emrich}, {Ewall-Wice}, {Feng}, {Gaensler}, {Goeke}, {Greenhill}, {Hazelton},
  {Jacobs}, {Kaplan}, {Kasper}, {Kratzenberg}, {Kudryavtseva}, {Lenc},
  {Lonsdale}, {Lynch}, {McWhirter}, {McKinley}, {Mitchell}, {Morales},
  {Morgan}, {Oberoi}, {Ord}, {Pindor}, {Prabu}, {Procopio}, {Offringa},
  {Riding}, {Rogers}, {Roshi}, {Shankar}, {Srivani}, {Subrahmanyan}, {Tingay},
  {Waterson}, {Wayth}, {Webster}, {Whitney}, {Williams}, \&
  {Williams}}]{hindson14}
{Hindson}, L., {Johnston-Hollitt}, M., {Hurley-Walker}, N., {et~al.} 2014,
  \mnras, 445, 330

\bibitem[{{Hoeft} \& {Br{\"u}ggen}(2007)}]{hoeft07}
{Hoeft}, M. \& {Br{\"u}ggen}, M. 2007, \mnras, 375, 77

\bibitem[{{Hoshino} {et~al.}(2010){Hoshino}, {Henry}, {Sato}, {Akamatsu},
  {Yokota}, {Sasaki}, {Ishisaki}, {Ohashi}, {Bautz}, {Fukazawa}, {Kawano},
  {Furuzawa}, {Hayashida}, {Tawa}, {Hughes}, {Kokubun}, \&
  {Tamura}}]{hoshino10}
{Hoshino}, A., {Henry}, J.~P., {Sato}, K., {et~al.} 2010, \pasj, 62, 371

\bibitem[{{Ibaraki} {et~al.}(2014){Ibaraki}, {Ota}, {Akamatsu}, {Zhang}, \&
  {Finoguenov}}]{ibaraki14}
{Ibaraki}, Y., {Ota}, N., {Akamatsu}, H., {Zhang}, Y.-Y., \& {Finoguenov}, A.
  2014, \aap, 562, A11

\bibitem[{{Ishisaki} {et~al.}(2007){Ishisaki}, {Maeda}, {Fujimoto}, {Ozaki},
  {Ebisawa}, {Takahashi}, {Ueda}, {Ogasaka}, {Ptak}, {Mukai}, {Hamaguchi},
  {Hirayama}, {Kotani}, {Kubo}, {Shibata}, {Ebara}, {Furuzawa}, {Iizuka},
  {Inoue}, {Mori}, {Okada}, {Yokoyama}, {Matsumoto}, {Nakajima}, {Yamaguchi},
  {Anabuki}, {Tawa}, {Nagai}, {Katsuda}, {Hayashida}, {Bamba}, {Miller},
  {Sato}, \& {Yamasaki}}]{ishisaki07}
{Ishisaki}, Y., {Maeda}, Y., {Fujimoto}, R., {et~al.} 2007, \pasj, 59, 113

\bibitem[{{Kaastra} {et~al.}(2009){Kaastra}, {Bykov}, \& {Werner}}]{kaastra09}
{Kaastra}, J.~S., {Bykov}, A.~M., \& {Werner}, N. 2009, \aap, 503, 373

\bibitem[{{Kang}(2015{\natexlab{a}})}]{kang15b}
{Kang}, H. 2015{\natexlab{a}}, Journal of Korean Astronomical Society, 48, 9

\bibitem[{{Kang}(2015{\natexlab{b}})}]{kang15a}
{Kang}, H. 2015{\natexlab{b}}, Journal of Korean Astronomical Society, 48, 155

\bibitem[{{Kang} {et~al.}(2005){Kang}, {Ryu}, {Cen}, \& {Song}}]{kang05}
{Kang}, H., {Ryu}, D., {Cen}, R., \& {Song}, D. 2005, \apj, 620, 21

\bibitem[{{Kang} {et~al.}(2012){Kang}, {Ryu}, \& {Jones}}]{kang12}
{Kang}, H., {Ryu}, D., \& {Jones}, T.~W. 2012, \apj, 756, 97

\bibitem[{{Kocevski} {et~al.}(2007){Kocevski}, {Ebeling}, {Mullis}, \&
  {Tully}}]{kocevski07}
{Kocevski}, D.~D., {Ebeling}, H., {Mullis}, C.~R., \& {Tully}, R.~B. 2007,
  \apj, 662, 224

\bibitem[{{Koyama} {et~al.}(1995){Koyama}, {Petre}, {Gotthelf}, {Hwang},
  {Matsuura}, {Ozaki}, \& {Holt}}]{koyama95}
{Koyama}, K., {Petre}, R., {Gotthelf}, E.~V., {et~al.} 1995, \nat, 378, 255

\bibitem[{{Kravtsov} \& {Borgani}(2012)}]{kravtsov12}
{Kravtsov}, A.~V. \& {Borgani}, S. 2012, \araa, 50, 353

\bibitem[{{Kushino} {et~al.}(2002){Kushino}, {Ishisaki}, {Morita}, {Yamasaki},
  {Ishida}, {Ohashi}, \& {Ueda}}]{kushino02}
{Kushino}, A., {Ishisaki}, Y., {Morita}, U., {et~al.} 2002, \pasj, 54, 327

\bibitem[{{Landau} \& {Lifshit's}(1959)}]{landau59}
{Landau}, L.~D. \& {Lifshit's}, E.~M. 1959, {Theory of elasticity}

\bibitem[{{Lodders}(2003)}]{lodders03}
{Lodders}, K. 2003, \apj, 591, 1220

\bibitem[{{Macario} {et~al.}(2011){Macario}, {Markevitch}, {Giacintucci},
  {Brunetti}, {Venturi}, \& {Murray}}]{macario11}
{Macario}, G., {Markevitch}, M., {Giacintucci}, S., {et~al.} 2011, \apj, 728,
  82

\bibitem[{{Markevitch}(2010)}]{markevitch10}
{Markevitch}, M. 2010, ArXiv e-prints

\bibitem[{{Markevitch} {et~al.}(2002){Markevitch}, {Gonzalez}, {David},
  {Vikhlinin}, {Murray}, {Forman}, {Jones}, \& {Tucker}}]{markevitch02}
{Markevitch}, M., {Gonzalez}, A.~H., {David}, L., {et~al.} 2002, \apjl, 567,
  L27

\bibitem[{{Markevitch} {et~al.}(2005){Markevitch}, {Govoni}, {Brunetti}, \&
  {Jerius}}]{markevitch05}
{Markevitch}, M., {Govoni}, F., {Brunetti}, G., \& {Jerius}, D. 2005, \apj,
  627, 733

\bibitem[{{Markevitch} \& {Vikhlinin}(2007)}]{markevitch07}
{Markevitch}, M. \& {Vikhlinin}, A. 2007, \physrep, 443, 1

\bibitem[{{Masai}(1984)}]{masai84}
{Masai}, K. 1984, \apss, 98, 367

\bibitem[{{Mazzotta} {et~al.}(2011){Mazzotta}, {Bourdin}, {Giacintucci},
  {Markevitch}, \& {Venturi}}]{mazzotta11}
{Mazzotta}, P., {Bourdin}, H., {Giacintucci}, S., {Markevitch}, M., \&
  {Venturi}, T. 2011, \memsai, 82, 495

\bibitem[{{Miniati} {et~al.}(2000){Miniati}, {Ryu}, {Kang}, {Jones}, {Cen}, \&
  {Ostriker}}]{miniati00}
{Miniati}, F., {Ryu}, D., {Kang}, H., {et~al.} 2000, \apj, 542, 608

\bibitem[{{Mitsuda} {et~al.}(2007){Mitsuda}, {Bautz}, {Inoue}, {Kelley},
  {Koyama}, {Kunieda}, {Makishima}, {Ogawara}, {Petre}, {Takahashi}, {Tsunemi},
  {White}, {Anabuki}, {Angelini}, {Arnaud}, {Awaki}, {Bamba}, {Boyce}, {Brown},
  {Chan}, {Cottam}, {Dotani}, {Doty}, {Ebisawa}, {Ezoe}, {Fabian}, {Figueroa},
  {Fujimoto}, {Fukazawa}, {Furusho}, {Furuzawa}, {Gendreau}, {Griffiths},
  {Haba}, {Hamaguchi}, {Harrus}, {Hasinger}, {Hatsukade}, {Hayashida}, {Henry},
  {Hiraga}, {Holt}, {Hornschemeier}, {Hughes}, {Hwang}, {Ishida}, {Ishisaki},
  {Isobe}, {Itoh}, {Iyomoto}, {Kahn}, {Kamae}, {Katagiri}, {Kataoka},
  {Katayama}, {Kawai}, {Kilbourne}, {Kinugasa}, {Kissel}, {Kitamoto}, {Kohama},
  {Kohmura}, {Kokubun}, {Kotani}, {Kotoku}, {Kubota}, {Madejski}, {Maeda},
  {Makino}, {Markowitz}, {Matsumoto}, {Matsumoto}, {Matsuoka}, {Matsushita},
  {McCammon}, {Mihara}, {Misaki}, {Miyata}, {Mizuno}, {Mori}, {Mori}, {Morii},
  {Moseley}, {Mukai}, {Murakami}, {Murakami}, {Mushotzky}, {Nagase}, {Namiki},
  {Negoro}, {Nakazawa}, {Nousek}, {Okajima}, {Ogasaka}, {Ohashi}, {Oshima},
  {Ota}, {Ozaki}, {Ozawa}, {Parmar}, {Pence}, {Porter}, {Reeves}, {Ricker},
  {Sakurai}, {Sanders}, {Senda}, {Serlemitsos}, {Shibata}, {Soong}, {Smith},
  {Suzuki}, {Szymkowiak}, {Takahashi}, {Tamagawa}, {Tamura}, {Tamura},
  {Tanaka}, {Tashiro}, {Tawara}, {Terada}, {Terashima}, {Tomida}, {Torii},
  {Tsuboi}, {Tsujimoto}, {Tsuru}, {Turner}, {Ueda}, {Ueno}, {Ueno}, {Uno},
  {Urata}, {Watanabe}, {Yamamoto}, {Yamaoka}, {Yamasaki}, {Yamashita},
  {Yamauchi}, {Yamauchi}, {Yaqoob}, {Yonetoku}, \& {Yoshida}}]{mitsuda07}
{Mitsuda}, K., {Bautz}, M., {Inoue}, H., {et~al.} 2007, \pasj, 59, 1

\bibitem[{{Molnar} {et~al.}(2013){Molnar}, {Chiu}, {Broadhurst}, \&
  {Stadel}}]{molnar13}
{Molnar}, S.~M., {Chiu}, I.-N.~T., {Broadhurst}, T., \& {Stadel}, J.~G. 2013,
  \apj, 779, 63

\bibitem[{{Nagai} \& {Lau}(2011)}]{nagai11}
{Nagai}, D. \& {Lau}, E.~T. 2011, \apjl, 731, L10

\bibitem[{{Nakazawa} {et~al.}(2009){Nakazawa}, {Sarazin}, {Kawaharada},
  {Kitaguchi}, {Okuyama}, {Makishima}, {Kawano}, {Fukazawa}, {Inoue},
  {Takizawa}, {Wik}, {Finoguenov}, \& {Clarke}}]{nakazawa09}
{Nakazawa}, K., {Sarazin}, C.~L., {Kawaharada}, M., {et~al.} 2009, \pasj, 61,
  339

\bibitem[{{Nandra} {et~al.}(2013){Nandra}, {Barret}, {Barcons}, {Fabian}, {den
  Herder}, {Piro}, {Watson}, {Adami}, {Aird}, {Afonso}, \& et~al.}]{nandra13}
{Nandra}, K., {Barret}, D., {Barcons}, X., {et~al.} 2013, ArXiv e-prints

\bibitem[{{Ogrean} \& {Br{\"u}ggen}(2013)}]{ogrean13_coma}
{Ogrean}, G.~A. \& {Br{\"u}ggen}, M. 2013, \mnras, 433, 1701

\bibitem[{{Ogrean} {et~al.}(2013){Ogrean}, {Br{\"u}ggen}, {R{\"o}ttgering},
  {Simionescu}, {Croston}, {van Weeren}, \& {Hoeft}}]{ogrean13}
{Ogrean}, G.~A., {Br{\"u}ggen}, M., {R{\"o}ttgering}, H., {et~al.} 2013,
  \mnras, 429, 2617

\bibitem[{{Ogrean} {et~al.}(2014){Ogrean}, {Br{\"u}ggen}, {van Weeren},
  {R{\"o}ttgering}, {Simionescu}, {Hoeft}, \& {Croston}}]{ogrean_chandra}
{Ogrean}, G.~A., {Br{\"u}ggen}, M., {van Weeren}, R., {et~al.} 2014, \mnras,
  440, 3416

\bibitem[{{Owers} {et~al.}(2014){Owers}, {Nulsen}, {Couch}, {Ma}, {David},
  {Forman}, {Hopkins}, {Jones}, \& {van Weeren}}]{owers14}
{Owers}, M.~S., {Nulsen}, P.~E.~J., {Couch}, W.~J., {et~al.} 2014, \apj, 780,
  163

\bibitem[{{Pinzke} {et~al.}(2013){Pinzke}, {Oh}, \& {Pfrommer}}]{pinzke13}
{Pinzke}, A., {Oh}, S.~P., \& {Pfrommer}, C. 2013, \mnras, 435, 1061

\bibitem[{{Reiprich} {et~al.}(2013){Reiprich}, {Basu}, {Ettori}, {Israel},
  {Lovisari}, {Molendi}, {Pointecouteau}, \& {Roncarelli}}]{reiprich13}
{Reiprich}, T.~H., {Basu}, K., {Ettori}, S., {et~al.} 2013, \ssr, 177, 195

\bibitem[{{Ricker} \& {Sarazin}(2001)}]{ricker01}
{Ricker}, P.~M. \& {Sarazin}, C.~L. 2001, \apj, 561, 621

\bibitem[{{Rudd} \& {Nagai}(2009)}]{rudd09}
{Rudd}, D.~H. \& {Nagai}, D. 2009, \apjl, 701, L16

\bibitem[{{Russell} {et~al.}(2012){Russell}, {McNamara}, {Sanders}, {Fabian},
  {Nulsen}, {Canning}, {Baum}, {Donahue}, {Edge}, {King}, \&
  {O'Dea}}]{russell12}
{Russell}, H.~R., {McNamara}, B.~R., {Sanders}, J.~S., {et~al.} 2012, \mnras,
  423, 236

\bibitem[{{Russell} {et~al.}(2010){Russell}, {Sanders}, {Fabian}, {Baum},
  {Donahue}, {Edge}, {McNamara}, \& {O'Dea}}]{russell10}
{Russell}, H.~R., {Sanders}, J.~S., {Fabian}, A.~C., {et~al.} 2010, \mnras,
  406, 1721

\bibitem[{{Ryu} {et~al.}(2003){Ryu}, {Kang}, {Hallman}, \& {Jones}}]{ryu03}
{Ryu}, D., {Kang}, H., {Hallman}, E., \& {Jones}, T.~W. 2003, \apj, 593, 599

\bibitem[{{Sarazin} {et~al.}(2014){Sarazin}, {Hogge}, {Chatzikos}, {Wik},
  {Giacintucci}, {Clarke}, {Wong}, {Gitti}, \& {Finoguenov}}]{sarazin14}
{Sarazin}, C., {Hogge}, T., {Chatzikos}, M., {et~al.} 2014, in The X-ray
  Universe 2014, edited by Jan-Uwe Ness. Online at <A
  href=''http://xmm.esac.esa.int/external/xmm\_science/workshops/2014symposium/''>http://xmm.esac.esa.int/external/xmm\_science/workshops/2014symposium/</A>,
  id.181

\bibitem[{{Sarazin}(1999)}]{sarazin99}
{Sarazin}, C.~L. 1999, \apj, 520, 529

\bibitem[{{Serlemitsos} {et~al.}(2007){Serlemitsos}, {Soong}, {Chan},
  {Okajima}, {Lehan}, {Maeda}, {Itoh}, {Mori}, {Iizuka}, {Itoh}, {Inoue},
  {Okada}, {Yokoyama}, {Itoh}, {Ebara}, {Nakamura}, {Suzuki}, {Ishida},
  {Hayakawa}, {Inoue}, {Okuma}, {Kubota}, {Suzuki}, {Osawa}, {Yamashita},
  {Kunieda}, {Tawara}, {Ogasaka}, {Furuzawa}, {Tamura}, {Shibata}, {Haba},
  {Naitou}, \& {Misaki}}]{xrt}
{Serlemitsos}, P.~J., {Soong}, Y., {Chan}, K.-W., {et~al.} 2007, \pasj, 59, 9

\bibitem[{{Simionescu} {et~al.}(2011){Simionescu}, {Allen}, {Mantz}, {Werner},
  {Takei}, {Morris}, {Fabian}, {Sanders}, {Nulsen}, {George}, \&
  {Taylor}}]{simionescu11}
{Simionescu}, A., {Allen}, S.~W., {Mantz}, A., {et~al.} 2011, Science, 331,
  1576

\bibitem[{{Skillman} {et~al.}(2013){Skillman}, {Xu}, {Hallman}, {O'Shea},
  {Burns}, {Li}, {Collins}, \& {Norman}}]{skillman13}
{Skillman}, S.~W., {Xu}, H., {Hallman}, E.~J., {et~al.} 2013, \apj, 765, 21

\bibitem[{{Smith} {et~al.}(2001){Smith}, {Brickhouse}, {Liedahl}, \&
  {Raymond}}]{apec}
{Smith}, R.~K., {Brickhouse}, N.~S., {Liedahl}, D.~A., \& {Raymond}, J.~C.
  2001, \apjl, 556, L91

\bibitem[{{Stroe} {et~al.}(2014{\natexlab{a}}){Stroe}, {Harwood}, {Hardcastle},
  \& {R{\"o}ttgering}}]{stroe14c}
{Stroe}, A., {Harwood}, J.~J., {Hardcastle}, M.~J., \& {R{\"o}ttgering},
  H.~J.~A. 2014{\natexlab{a}}, \mnras, 445, 1213

\bibitem[{{Stroe} {et~al.}(2014{\natexlab{b}}){Stroe}, {Rumsey}, {Harwood},
  {van Weeren}, {R{\"o}ttgering}, {Saunders}, {Sobral}, {Perrott}, \&
  {Schammel}}]{stroe14b}
{Stroe}, A., {Rumsey}, C., {Harwood}, J.~J., {et~al.} 2014{\natexlab{b}},
  \mnras, 441, L41

\bibitem[{{Stroe} {et~al.}(2013){Stroe}, {van Weeren}, {Intema},
  {R{\"o}ttgering}, {Br{\"u}ggen}, \& {Hoeft}}]{stroe13}
{Stroe}, A., {van Weeren}, R.~J., {Intema}, H.~T., {et~al.} 2013, \aap, 555,
  A110

\bibitem[{{Takahashi} {et~al.}(2012){Takahashi}, {Mitsuda}, {Kelley}, {Aarts},
  {Aharonian}, {Akamatsu}, {Akimoto}, {Allen}, {Anabuki}, {Angelini}, {Arnaud},
  {Asai}, {Audard}, {Awaki}, {Azzarello}, {Baluta}, {Bamba}, {Bando}, {Bautz},
  {Blandford}, {Boyce}, {Brown}, {Cackett}, {Chernyakova}, {Coppi},
  {Costantini}, {de Plaa}, {den Herder}, {DiPirro}, {Done}, {Dotani}, {Doty},
  {Ebisawa}, {Eckart}, {Enoto}, {Ezoe}, {Fabian}, {Ferrigno}, {Foster},
  {Fujimoto}, {Fukazawa}, {Funk}, {Furuzawa}, {Galeazzi}, {Gallo}, {Gandhi},
  {Gendreau}, {Gilmore}, {Haas}, {Haba}, {Hamaguchi}, {Hatsukade}, {Hayashi},
  {Hayashida}, {Hiraga}, {Hirose}, {Hornschemeier}, {Hoshino}, {Hughes},
  {Hwang}, {Iizuka}, {Inoue}, {Ishibashi}, {Ishida}, {Ishimura}, {Ishisaki},
  {Ito}, {Iwata}, {Iyomoto}, {Kaastra}, {Kallman}, {Kamae}, {Kataoka},
  {Katsuda}, {Kawahara}, {Kawaharada}, {Kawai}, {Kawasaki}, {Khangaluyan},
  {Kilbourne}, {Kimura}, {Kinugasa}, {Kitamoto}, {Kitayama}, {Kohmura},
  {Kokubun}, {Kosaka}, {Koujelev}, {Koyama}, {Krimm}, {Kubota}, {Kunieda},
  {LaMassa}, {Laurent}, {Lebrun}, {Leutenegger}, {Limousin}, {Loewenstein},
  {Long}, {Lumb}, {Madejski}, {Maeda}, {Makishima}, {Marchand}, {Markevitch},
  {Matsumoto}, {Matsushita}, {McCammon}, {McNamara}, {Miller}, {Miller},
  {Mineshige}, {Minesugi}, {Mitsuishi}, {Miyazawa}, {Mizuno}, {Mori}, {Mori},
  {Mukai}, {Murakami}, {Murakami}, {Mushotzky}, {Nagano}, {Nagino}, {Nakagawa},
  {Nakajima}, {Nakamori}, {Nakazawa}, {Namba}, {Natsukari}, {Nishioka},
  {Nobukawa}, {Nomachi}, {O'Dell}, {Odaka}, {Ogawa}, {Ogawa}, {Ogi}, {Ohashi},
  {Ohno}, {Ohta}, {Okajima}, {Okamoto}, {Okazaki}, {Ota}, {Ozaki}, {Paerels},
  {Paltani}, {Parmar}, {Petre}, {Pohl}, {Porter}, {Ramsey}, {Reis}, {Reynolds},
  {Russell}, {Safi-Harb}, {Sakai}, {Sameshima}, {Sanders}, {Sato}, {Sato},
  {Sato}, {Sato}, {Sawada}, {Serlemitsos}, {Seta}, {Shibano}, {Shida},
  {Shimada}, {Shinozaki}, {Shirron}, {Simionescu}, {Simmons}, {Smith},
  {Sneiderman}, {Soong}, {Stawarz}, {Sugawara}, {Sugita}, {Sugita},
  {Szymkowiak}, {Tajima}, {Takahashi}, {Takeda}, {Takei}, {Tamagawa}, {Tamura},
  {Tamura}, {Tanaka}, {Tanaka}, {Tashiro}, {Tawara}, {Terada}, {Terashima},
  {Tombesi}, {Tomida}, {Tsuboi}, {Tsujimoto}, {Tsunemi}, {Tsuru}, {Uchida},
  {Uchiyama}, {Uchiyama}, {Ueda}, {Ueno}, {Uno}, {Urry}, {Ursino}, {de Vries},
  {Wada}, {Watanabe}, {Werner}, {White}, {Yamada}, {Yamada}, {Yamaguchi},
  {Yamasaki}, {Yamauchi}, {Yamauchi}, {Yatsu}, {Yonetoku}, {Yoshida}, \&
  {Yuasa}}]{takahashi12}
{Takahashi}, T., {Mitsuda}, K., {Kelley}, R., {et~al.} 2012, in Society of
  Photo-Optical Instrumentation Engineers (SPIE) Conference Series, Vol. 8443,
  Society of Photo-Optical Instrumentation Engineers (SPIE) Conference Series

\bibitem[{{Takizawa}(2005)}]{takizawa05}
{Takizawa}, M. 2005, \apj, 629, 791

\bibitem[{{Takizawa} \& {Mineshige}(1998)}]{takizawa98}
{Takizawa}, M. \& {Mineshige}, S. 1998, \apj, 499, 82

\bibitem[{{Tawa} {et~al.}(2008){Tawa}, {Hayashida}, {Nagai}, {Nakamoto},
  {Tsunemi}, {Yamaguchi}, {Ishisaki}, {Miller}, {Mizuno}, {Dotani}, {Ozaki}, \&
  {Katayama}}]{tawa08}
{Tawa}, N., {Hayashida}, K., {Nagai}, M., {et~al.} 2008, \pasj, 60, 11

\bibitem[{{van Adelsberg} {et~al.}(2008){van Adelsberg}, {Heng}, {McCray}, \&
  {Raymond}}]{vanadelsberg08}
{van Adelsberg}, M., {Heng}, K., {McCray}, R., \& {Raymond}, J.~C. 2008, \apj,
  689, 1089

\bibitem[{{van Weeren} {et~al.}(2011){van Weeren}, {Br{\"u}ggen},
  {R{\"o}ttgering}, \& {Hoeft}}]{vanweeren11_sim}
{van Weeren}, R.~J., {Br{\"u}ggen}, M., {R{\"o}ttgering}, H.~J.~A., \& {Hoeft},
  M. 2011, \mnras, 418, 230

\bibitem[{{van Weeren} {et~al.}(2010){van Weeren}, {R{\"o}ttgering},
  {Br{\"u}ggen}, \& {Hoeft}}]{vanweeren10}
{van Weeren}, R.~J., {R{\"o}ttgering}, H.~J.~A., {Br{\"u}ggen}, M., \& {Hoeft},
  M. 2010, Science, 330, 347

\bibitem[{{van Weeren} {et~al.}(2012){van Weeren}, {R{\"o}ttgering}, {Intema},
  {Rudnick}, {Br{\"u}ggen}, {Hoeft}, \& {Oonk}}]{vanweeren12_toothbrush}
{van Weeren}, R.~J., {R{\"o}ttgering}, H.~J.~A., {Intema}, H.~T., {et~al.}
  2012, \aap, 546, A124

\bibitem[{{Voit}(2005)}]{voit05}
{Voit}, G.~M. 2005, Reviews of Modern Physics, 77, 207

\bibitem[{{Werner} {et~al.}(2013){Werner}, {Urban}, {Simionescu}, \&
  {Allen}}]{werner13}
{Werner}, N., {Urban}, O., {Simionescu}, A., \& {Allen}, S.~W. 2013, \nat, 502,
  656

\bibitem[{{Willingale} {et~al.}(2013){Willingale}, {Starling}, {Beardmore},
  {Tanvir}, \& {O'Brien}}]{willingale13}
{Willingale}, R., {Starling}, R.~L.~C., {Beardmore}, A.~P., {Tanvir}, N.~R., \&
  {O'Brien}, P.~T. 2013, \mnras, 431, 394

\bibitem[{{Yoshino} {et~al.}(2009){Yoshino}, {Mitsuda}, {Yamasaki}, {Takei},
  {Hagihara}, {Masui}, {Bauer}, {McCammon}, {Fujimoto}, {Wang}, \&
  {Yao}}]{yoshino09}
{Yoshino}, T., {Mitsuda}, K., {Yamasaki}, N.~Y., {et~al.} 2009, \pasj, 61, 805

\end{thebibliography}

\end{document}